\documentclass[
aps,
prd,
onecolumn,
superscriptaddress,
floatfix,
longbibliography,
nofootinbib,
preprintnumbers,
10pt
]{revtex4-2}
\usepackage{graphicx}
\usepackage{dcolumn}
\usepackage{bm}
\usepackage{ulem}
\usepackage{amsmath}
\usepackage{amssymb}
\usepackage{txfonts}
\usepackage{color} 
\usepackage{xcolor}
\usepackage{listings}
\usepackage{cprotect}
\usepackage{hyperref}
\usepackage{tikz}
\usetikzlibrary{quantikz2}

\graphicspath{{./figures/}}

\newcommand{\Tr}{{\rm Tr}}
\newcommand{\e}{{\rm e}}
\newcommand{\imag}{{\rm i}}

\newcommand{\im}{\mathrm{i}}

\newcommand{\calD}{\mathcal{D}}

\newcommand{\calO}{\mathcal{O}}
\newcommand{\calU}{\mathcal{U}}

\hypersetup{
  colorlinks=true,
  linkcolor=[rgb]{0.60,0.00,0.00},
  citecolor=[rgb]{0.00,0.00,0.60},
  urlcolor=[rgb]{0.00,0.00,0.60},
  setpagesize=false
}

\definecolor{codegreen}{rgb}{0,0.6,0}
\definecolor{codegray}{rgb}{0.5,0.5,0.5}
\definecolor{codepurple}{rgb}{0.58,0,0.82}
\definecolor{backcolour}{rgb}{0.95,0.95,0.92}

\lstdefinestyle{mystyle}{
    backgroundcolor=\color{backcolour},   
    commentstyle=\color{codegreen},
    keywordstyle=\color{magenta},
    numberstyle=\tiny\color{codegray},
    stringstyle=\color{codepurple},
    basicstyle=\ttfamily\footnotesize,
    breakatwhitespace=false,         
    breaklines=true,                 
    captionpos=b,                    
    keepspaces=true,                 
    numbers=left,                    
    numbersep=5pt,                  
    showspaces=false,                
    showstringspaces=false,
    showtabs=false,                  
    tabsize=2
}

\begin{document}

\title{
  Floquet evolution of the q-deformed \texorpdfstring{SU(3)${}_1$}{SU(3)1} Yang-Mills theory on a two-leg ladder
}

\author{Tomoya Hayata}
\email{hayata@keio.jp}
\affiliation{Departments of Physics, Keio University School of Medicine, 4-1-1 Hiyoshi, Kanagawa 223-8521, Japan}
\affiliation{Interdisciplinary Theoretical and Mathematical Sciences Program (iTHEMS), RIKEN, Wako, Saitama 351-0198, Japan}

\author{Yoshimasa Hidaka}
\email{yoshimasa.hidaka@yukawa.kyoto-u.ac.jp}
\affiliation{Yukawa Institute for Theoretical Physics, Kyoto University, Kyoto 606-8502, Japan}
\affiliation{Interdisciplinary Theoretical and Mathematical Sciences Program (iTHEMS), RIKEN, Wako, Saitama 351-0198, Japan}

\preprint{RIKEN-iTHEMS-Report-24, YITP-24-116}

\begin{abstract}

We simulate Floquet time-evolution of a truncated SU(3) lattice Yang-Mills theory on a two-leg ladder geometry under open boundary conditions using IBM's superconducting 156-qubit device ibm\_fez. To this end, we derive the quantum spin representation of the lattice Yang-Mills theory, and compose a quantum circuit carefully tailored to hard wares, reducing the use of CZ gates. 
Since it is still challenging to simulate Hamiltonian evolution in present noisy quantum processors, we make the step size in the Suzuki-Trotter decomposition very large, and simulate thermalization dynamics in Floquet circuit composed of the Suzuki-Trotter evolution. 
We demonstrate that IBM's Heron quantum processor can simulate, by error mitigation, Floqeut thermalization dynamics in a large system consisting of $62$ qubits. Our work would be a benchmark for further quantum simulations of lattice gauge theories using real devices.
 
\end{abstract}

\date{\today}
\maketitle

\section{Introduction}
\label{sec:introduction}
Studying the potential power of quantum computers has been one of the central subjects in high-energy physics.
In particular, digital quantum simulation of lattice gauge theories has been intensively discussed, which would be potentially useful e.g., to study real-time evolution of QCD for understanding the nonequilibirum dynamics of quark-gluon-plasmas in heavy-ion collision experiments, or to compute the equation of state of QCD at finite density, which is important to understand the physics of neutron stars. Both cannot be solved by conventional Monte Carlo methods due to the notorious sign problem.

There has been a considerable progress in Hamiltonian formulation of lattice gauge theories, and some benchmark studies to test the potential of quantum computers were performed using $(1+1)$- and $(2+1)$-dimensional lattice gauge theories~\cite{Klco:2018kyo,Klco:2019evd,Ciavarella:2021nmj,Atas:2021ext,deJong:2021wsd,Ciavarella:2021lel,Nguyen:2021hyk,Atas:2022dqm,ARahman:2022tkr,Charles:2023zbl,Farrell:2023fgd,Schuster:2023klj,Angelides:2023noe,Farrell:2024fit,Ciavarella:2024fzw,Hayata:2024smx,Cochran:2024rwe}. Since quantum simulations were successful in running complex circuits consisting of $\calO(50-100)$ qubits in those benchmark studies, one may expect that present quantum computers have enough capability to solve some important problems in lattice gauge theories. However, it is believed that simulating Hamiltonian dynamics of quantum many-body systems including lattice gauge theories in physically relevant setup requires the realization of fault-tolerant quantum computers. Present quantum computations in any architecture are necessarily subject to noise. This severely limits the capability of current quantum computers, so that we are able to simulate only very short timescales even taking error mitigation into consideration. 

Under such circumstances, the nonequilibrium dynamics of the Floquet system with classically intractable system sizes  has attracted increasing attention as a physical problem that might be feasible in near-term quantum devices~\cite{Eckstein:2023sjk,Yang:2023nak,Shinjo:2024vci,Seki:2024rfx,Hayata:2024smx}. Indeed, we often make a time step $\delta t$ in the Suzuki-Trotter decomposition of Hamiltonian evolution operator $e^{-i\delta t H}$ as large as possible within acceptable accuracy to reduce the depth of circuits. Once we abandon the goal of faithfully simulating Hamiltonian evolution, we can make $\delta t$ much larger than that used in Hamiltonian evolution, and obtain a quantum circuit feasible to study in near-term quantum devices. Such a quantum circuit can be understood as a Floquet system, whose driving period is nothing but one time step of the Suzuki-Trotter formula.
Even though Floquet systems heat up to infinite temperature, it is known that the heating time can be exponentially long by controlling driving frequency, i.e., the inverse of the time step, and Floquet system has a long-lived prethermal state~\cite{Lazarides2014,DAlessio2014,Abanin2015,Mori2016,Kuwahara2016,Mori:2017qhg}. It was proposed to utilize the prethermal state for studying thermal properties of quantum many-body systems using near-term digital quantum computers~\cite{Yang:2023nak}. We also utilize the Floquet circuit to study the thermalization dynamics of a lattice gauge theory using present noisy quantum computers.

In this paper, we simulate thermalization dynamics in the Floquet time-evolution of a truncated SU(3) lattice Yang-Mills theory on a two-leg ladder geometry under open boundary conditions using IBM's Heron quantum processor ibm\_fez. We consider SU(3) to study a more complex many-body system than SU(2)~\cite{Klco:2019evd,Hayata:2021kcp,Ciavarella:2021nmj,Ciavarella:2021lel}, and test the capabilities of quantum computers. We truncate the irreducible representations of the SU($3$) group to its lowest nontrivial dimension of the Dynkin index, and employ the $q$-deformed formulation of the Yang-Mills theory~\cite{Zache:2023dko,Hayata:2023puo,Hayata:2023bgh}, where the truncated SU($3$) group is described as the SU(3)${}_1$ quantum group. We derive the explicit quantum spin representation of the model, and compose a quantum circuit carefully tailored to hardwares, reducing the use of CZ gates. We demonstrate that even a present noisy quantum device can simulate, with the help of error mitigation, Floquet evolution in a large system consisting of $62$ qubits.

The rest of this paper is organized as follows. In Sec.~\ref{sec:model}, we introduce the $q$-deformed SU(3)${}_1$ Yang-Mills theory on a two-leg ladder geometry. We derive the explicit spin representation of the Hamiltonian of the model. We show that the model is described as a spin-one quantum chain, i.e., a one-dimensional qutrit system. In Sec.~\ref{sec:method}, we explain the quantum circuit to simulate the Floquet dynamics of the $q$-deformed SU(3)${}_1$ Yang-Mills theory. Since IBM's superconducting quantum processor is composed of qubits, we need to map the qutrit chain obtained in Sec.~\ref{sec:model} to qubits. The details of mapping are given in Sec.~\ref{sec:method}.
The experimental setups, raw data, and results of error mitigation are presented in Sec.~\ref{sec:experiments}.
Finally, we give a discussion in Sec.~\ref{sec:conclusion}. 

\section{Model}
\label{sec:model}

We consider the $q$-deformed SU($3$) Yang-Mills theory on a two-leg ladder geometry under open boundary conditions. We truncate irreducible representations of the SU($3$) group by its lowest nontrivial dimension of the Dynkin index to simplify the model as much as possible. Furthermore, to keep the unitarity of Wilson loops manifestly, we employ the $q$ deformation of the truncated SU($3$) that we call SU($3$)$_1$. We note that we need to consider SU($3$)$_k$ symmetry with various integer $k$, and extrapolate the $k$ dependence of observables to $k\rightarrow\infty$ for simulating the real SU($3$) Yang-Mills theory (see, e.g., Ref.~\cite{Hayata:2023bgh}), but it is not feasible in present quantum computers. We also consider a two-leg ladder geometry as a minimal lattice in which the plaquette operators can be defined. 

The fusion rules of SU($3$)$_1$, i.e., the composition rules of Wilson lines, are the same as those of $\mathbb{Z}_3$ group, and are given explicitly as
\begin{align}
\label{eq:fusionrules}
    1\times1=1,\;\;1\times3=3\times1=3,\;\;1\times\bar{3}=\bar{3}\times1=\bar{3},
    \notag \\
    3\times 3=\bar{3},\;\;3\times\bar{3}=1,\;\;\bar{3}\times\bar{3}=3,
\end{align}
where $1\equiv(0,0)$, $3\equiv(1,0)$, and $\bar{3}\equiv(0,1)$ represent the Dynkin indices of SU($3$) labeled by its the dimension. 
We note that $\bar{\bar{3}}=3$, and SU($3$)$_1$ is multiplicity free~\cite{Hayata:2023bgh}.
We can label the computational basis of the Hilbert space by using the Dynkin indices and directed arrows. Indeed, such a basis is known as the string network. As an example, we show a string network of the four plaquette system:
\begin{equation}
\label{eq:stringnet}
  \parbox{8.cm}{\includegraphics[scale=0.3]{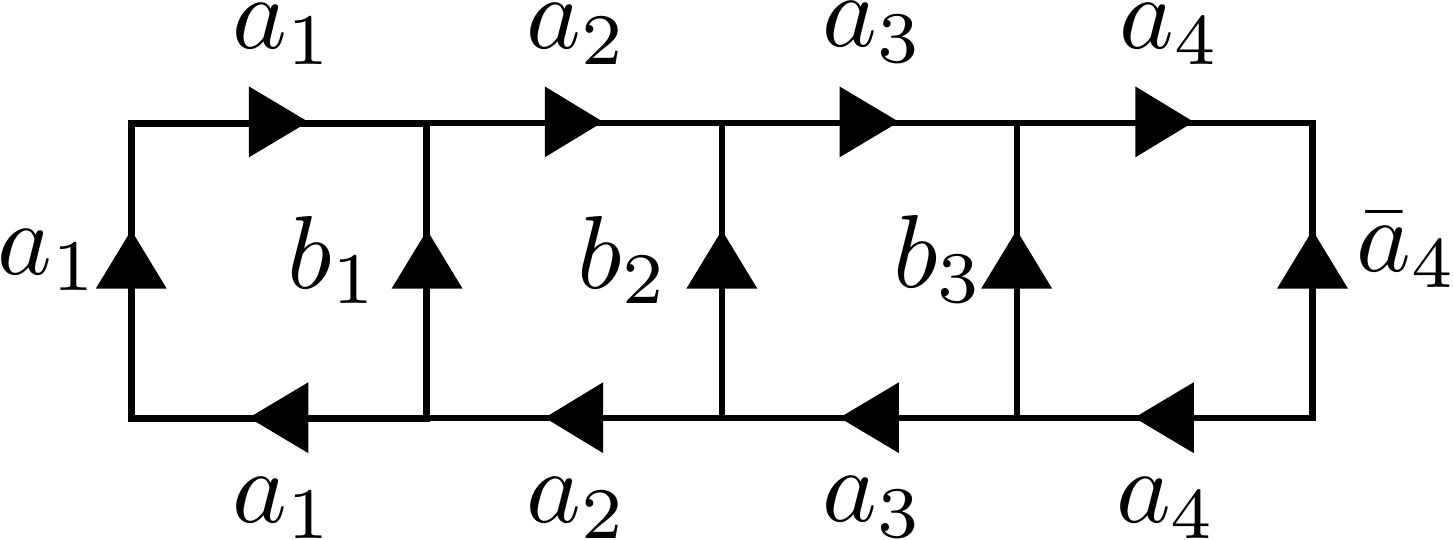}} ,
\end{equation}
where $a_i$ and $b_i$ are the Dynkin indices, respectively.
$a_i$ and $b_i$ that meet at vertices satisfy the fusion rules~\eqref{eq:fusionrules}, so that $b_i$ on the rungs are completely fixed by $a_i$ and $a_{i+1}$. 
Therefore, this system can be labeled by a chain of $a_i$.
For example, the quantum state of the system in Eq.~\eqref{eq:stringnet} is written as 
\begin{equation}
|\psi\rangle=\sum_{a_1,a_2,a_3,a_4\in\{1,3,\bar{3}\}}\psi(a_1,a_2,a_3,a_4)|a_1,a_2,a_3,a_4\rangle .
\end{equation}
The dimension of the Hilbert space is scaled as $3^N$ with $N$ being the number of plaquettes, and thus the system can be described as a $S=1$ quantum spin chain i.e., a qutrit chain.

In what follows, we derive the explicit spin representation of the $q$-deformed SU($3$)$_1$ Kogut-Susskind Hamiltonian to implement it on quantum processors. To this end, we follows Ref.~\cite{Gils2009}, and move to the unitary equivalent basis by applying $F$-moves to rungs.
Here, the $F$-moves are defined as
\begin{align}
\label{eq:wilsonloop_fusion_graph}
  \parbox{2.cm}{\includegraphics[scale=0.3]{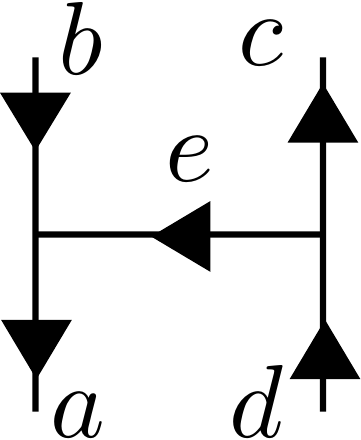}}
  =\sum_f[F^{a\bar{b}c}_d]_{ef}\parbox{2.cm}{\includegraphics[scale=0.3]{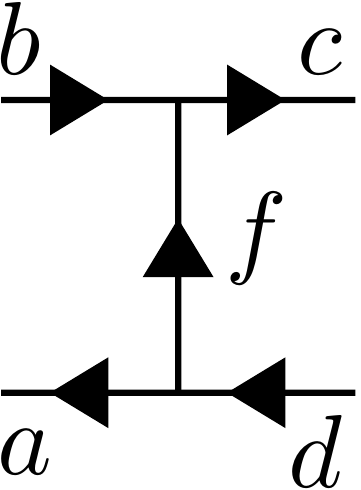}}.
\end{align}
A new basis can be labeled as 
\begin{equation}
\label{eq:tadpole}
  \parbox{8.cm}{\includegraphics[scale=0.3]{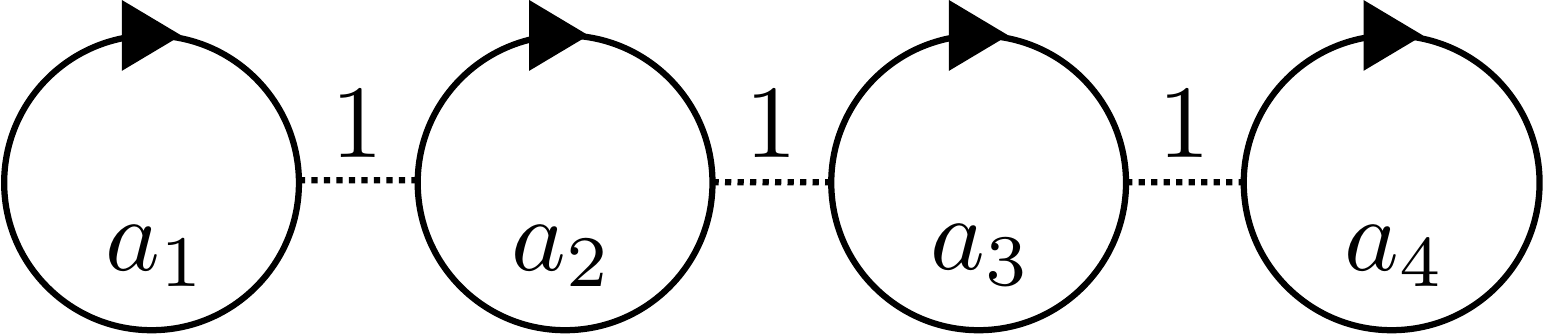}} .
\end{equation}
In this basis, the rungs are not dynamical and can be removed explicitly.
Since plaquettes are isolated, the action of the Wilson loop $c$ (operator) to the string $a$ (state) is determined solely by fusion rules $c\times a=N_{ca}^b b$ in Eq.~\eqref{eq:fusionrules} as $c|a\rangle=\sum_b N_{ca}^b |b\rangle$.
Thus, the fundamental plaquette operator reads
\begin{equation}
    W=
    \begin{pmatrix}
    0 & 0 & 1 \\
    1 & 0 & 0 \\
    0 & 1 & 0 
    \end{pmatrix} ,
\end{equation}
where we identified $|1\rangle$, $|3\rangle$, and $|\bar{3}\rangle$ as $(1,0,0)$, $(0,1,0)$, and $(0,0,1)$, respectively.
Putting the $S=1$ spins on the center of plaquettes, the magnetic part of the Kogut-Suskkind Hamiltonian of the $N$ plaquette system reads
\begin{equation}
    \label{eq:magnetic}
    H_B=-\frac{K}{2}\sum_{x=1}^N\bigl[ W(x)+W^\dagger(x) \bigr],
\end{equation}
where $K=1/g^4$ is the coupling constant in lattice unit, and $x$ labels the position of the $S=1$ spin.

Next, we consider the electric part of the Hamiltonian.
The $E^2$ operator is diagonal in the original string-net basis~\eqref{eq:stringnet}, and is given explicitly as 
\begin{equation}
    E^2=\sum_{a\in\{1,3,\bar{3}\}} C(a)|a\rangle\langle a|,
\end{equation}
where $C(a)$ is the Casimir invariant~\cite{10.1063/1.529497} 
\begin{equation}
C(a)=\begin{cases}
0 \;\;(a=1)\\
c \;\;(a=3,\bar{3})
    \end{cases} ,
\end{equation}
with $c=3/4$.
The $E^2$ operator acts on all edges ($a_i$ and $b_i$) in the string-net basis~\eqref{eq:stringnet}.
Since the strings defined on two-legs do not change under the unitary transformation by $F$-moves, the action of $E^2$ operator on the legs are also invariant.
We have two contributions. One is the $E^2$ operator directly acting on the legs, which reads
\begin{align}
    \label{eq:casimir}
    M=&2\sum_{a=3,\bar{3}}|a\rangle\langle a| ,
\end{align}
at the middle of a chain, and 
\begin{align}
    \label{eq:casimir2}
    M=&3\sum_{a=3,\bar{3}}|a\rangle\langle a| ,
\end{align}
at the ends of a chain since the plaquette has three edges as understood from Eq.~\eqref{eq:stringnet}. 
The other is the $E^2$ operator acting on rungs ($b_i$ in the original string-net basis), which gives two-body operators in terms of $a_i$ and reads
\begin{align}
    \label{eq:casimir3}
    M^r=&\sum_{a=3,\bar{3}}\bigl(|a,1\rangle\langle a,1|+|1,a\rangle\langle 1,a|+|a,\bar{a}\rangle\langle a,\bar{a}|\bigr) .
\end{align}
Therefore, the electric part of the Hamiltonian is
\begin{equation}
    \label{eq:electric}
    H_E=\frac{c}{2}\sum_{x=1}^N M(x)+\frac{c}{2}\sum_{x=1}^{N-1}M^r(x,x+1)\;\;.
\end{equation}
Finally, the spin representation of the Kogut-Susskind Hamiltonian is the sum of electric and magnetic parts: $H=H_E+H_B$.


\section{Method}
\label{sec:method}
We aim to simulate the time evolution of the $S=1$ spin chain on a superconducting quantum device.
To this end, we employ two qubits to describe a plaquette (i.e., a qutrit).
The explicit mapping is given as $|1\rangle\rightarrow |00\rangle$, $|3\rangle\rightarrow |01\rangle$, and $|\bar{3}\rangle\rightarrow |10\rangle$.
Namely, we put a pair of qubits in the center of the plaquettes.
We define the projectors acting on two qubits as
\begin{eqnarray}
    M_1 &=& \frac{1+Z}{2}\otimes\frac{1+Z}{2} ,
    \\
    M_{3} &=&\frac{1+Z}{2}\otimes\frac{1-Z}{2} ,
    \\
    M_{\bar{3}} &=&\frac{1-Z}{2}\otimes\frac{1+Z}{2} ,
    \\
    M_{\phi} &=&\frac{1-Z}{2}\otimes\frac{1-Z}{2} .
\end{eqnarray}
Using them, the Casimir~\eqref{eq:casimir} is written as
\begin{align}
    \label{eq:casimir_qubit}
    M^{(q)}&=2M_3+2M_{\bar{3}}
    \notag    \\
    &=1-Z_1Z_2 ,
\end{align}
where we label two qubits that describe a qutrit as $1$ and $2$. Similarly, the Casimir in Eq.~\eqref{eq:casimir3} is written as
\begin{align}
    M^{r(q)}&=\frac{1}{2}M\otimes M_{1}+\frac{1}{2}M_1\otimes M
    +M_3\otimes M_{\bar{3}}
    +M_{\bar{3}}\otimes M_3 ,
\end{align}
which is expressed explicitly as
\begin{align}
    M^{r(q)} &=\frac{1}{8}(3+Z_1+Z_2+Z_3+Z_4)
    \notag \\
    &\quad+\frac{1}{8} (-Z_1Z_2+Z_2Z_3-Z_3Z_4+Z_1Z_4-Z_1Z_3-Z_2Z_4)
    \notag \\
    &\quad+\frac{1}{8} (-Z_1Z_2Z_3-Z_2Z_3Z_4-Z_1Z_2Z_4-Z_1Z_3Z_4)
    \notag \\
    &\quad-\frac{1}{8}Z_1Z_2Z_3Z_4 ,
\end{align}
where we label four qubits that describe two qutrits as $1$, $2$, $3$ and $4$, respectively. As easily understood from the above expression, implementation of the rung operator is very challenging in present noisy quantum processors. To reduce the number of two-qubit gates, we add terms such that they do not change the dynamics inside the logical Hilbert space spanned by $|1\rangle\rightarrow |00\rangle$, $|3\rangle\rightarrow |01\rangle$, and $|\bar{3}\rangle\rightarrow |10\rangle$, and can remove multi-qubit interactions. We modify the rung operator as
\begin{align}
\label{eq:rung}
    \bar{M}^{r(q)} &=M^{r(q)}-2M_{\phi}\otimes M_{\phi}+2M_{1}\otimes M_{\phi}+2M_{\phi}\otimes M_{1}
    \notag \\
    &=\frac{1}{2}+\frac{1}{4} (Z_1+Z_2+Z_3+Z_4)
    \notag \\
    &-\frac{1}{2} (Z_1Z_3+Z_2Z_4)-\frac{1}{4} (Z_2Z_3+Z_1Z_4).
\end{align}
The additional terms do not affect the logical qutrit state $|\Psi_{\rm trit}\rangle$ since they satisfy $M_{\phi}|\Psi_{\rm trit}\rangle=0$.
Unfortunately, there are still a few nonlocal CZ gates. For the implementation of the remaining nonlocal CZ gates in Eq.~\eqref{eq:rung}, we rely on optimization of circuits using Qiskit's transpiler~\cite{qiskit_paper}.

Next, we explain the qubit representation of the plaquette operator. We implement the plaquette operator so that it is still a unitary operator in an enlarged Hilbert space. The fundamental Wilson loop operator reads
\begin{align}
    W^{(q)}=
    &\begin{pmatrix}
    0 & 0 & 1 & 0\\
    1 & 0 & 0 & 0\\
    0 & 1 & 0 & 0 \\
    0 & 0 & 0 & 1
    \end{pmatrix}  ,
\end{align}
where we identified $|00\rangle$, $|01\rangle$, $|10\rangle$ and $|11\rangle$ as $(1,0,0,0)$, $(0,1,0,0)$, $(0,0,1,0)$ and $(0,0,0,1)$, respectively. Since $W$ is a unitary operator, we can write it as $W=Q\Lambda Q^\dag$ where 
\begin{align}
    Q=
    &\begin{pmatrix}
    \frac{1}{\sqrt{3}} & -\frac{1}{2\sqrt{3}}-\frac{\im}{2} &  -\frac{1}{2\sqrt{3}}+\frac{\im}{2} & 0\\
    \frac{1}{\sqrt{3}} & -\frac{1}{2\sqrt{3}}+\frac{\im}{2} & -\frac{1}{2\sqrt{3}}-\frac{\im}{2} & 0\\
    \frac{1}{\sqrt{3}} & \frac{1}{\sqrt{3}} & \frac{1}{\sqrt{3}} & 0 \\
    0 & 0 & 0 & 1
    \end{pmatrix} ,
\end{align}
and
\begin{align}
\label{eq:observable}
    \Lambda=
    &\begin{pmatrix}
    1 & 0 &  0 & 0\\
    0 & e^{\frac{2\im\pi}{3}} & 0 & 0\\
    0 &0 & e^{-\frac{2\im\pi}{3}} & 0 \\
    0 & 0 & 0 & 1
    \end{pmatrix} .
\end{align}
Those act as two-qubit operators in real qubit devices. We note that we use Qiskit's unitary gate to implement $Q$ in the following experiments.

\begin{figure}
    \begin{quantikz}[column sep=.5cm, row sep=.1cm]
         & \gate[style={fill=red!20}]{} 
         & \gate[style={fill=green!20}]{} 
         & \gate[2, style={fill=green!20}]{}
         & 
         & 
         \\
         & \gate[style={fill=red!20}]{} 
         & \gate[style={fill=green!20}]{}
         & 
         & \gate[2, style={fill=green!20}]{}
         & 
         \\
         & \gate[style={fill=red!20}]{}  
         & \gate[style={fill=green!20}]{}
         & \gate[2, style={fill=green!20}]{}
         & 
         & 
         \\
         & \gate[style={fill=red!20}]{} 
         & \gate[style={fill=green!20}]{} 
         & 
         & \gate[2, style={fill=green!20}]{}
         & 
         \\
         & \gate[style={fill=red!20}]{} 
         & \gate[style={fill=green!20}]{}
         & \gate[2, style={fill=green!20}]{}
         & 
         & 
         \\
         & \gate[style={fill=red!20}]{} 
         & \gate[style={fill=green!20}]{}
         & 
         & \gate[2, style={fill=green!20}]{}
         & 
         \\
         & \gate[style={fill=red!20}]{} 
         & \gate[style={fill=green!20}]{}
         & \gate[2, style={fill=green!20}]{}
         & 
         & 
         \\
         & \gate[style={fill=red!20}]{}  
         & \gate[style={fill=green!20}]{}
         & 
         & \gate[2, style={fill=green!20}]{}
         & 
         \\
         & \gate[style={fill=red!20}]{} 
         & \gate[style={fill=green!20}]{} 
         & 
         & 
         & 
    \end{quantikz}
    \\
    \begin{quantikz}[column sep=.3cm, row sep=.1cm]
    &\gate[style={fill=red!20}]{}&
    \end{quantikz}$ 
    = Q e^{\im \frac{K}{2}dt \left(\Lambda+\Lambda^\dagger\right)}Q^\dagger$
    \quad
    \begin{quantikz}[column sep=.3cm, row sep=.1cm]
    &\gate[style={fill=green!20}]{}&
    \end{quantikz}$ 
    = \e^{-\imag \frac{dt}{2} M^{(q)}}$
    \quad
    \begin{quantikz}[column sep=.3cm, row sep=.1cm]
    &\gate[2, style={fill=green!20}]{}&\\ &&
    \end{quantikz}$ 
    = \e^{-\im\frac{dt}{2} \bar{M}^{r(q)}}$
    
    \caption{Circuit diagram of a single Trotter step $\calU(T)$ for $N=9$ plaquette system under open boundary conditions. Here, each line represents a qutrit that is simulated by using two qubits, and the boxes act as two- and four-qubit gates in real devices.
    \label{fig:UFcirc}
    }
    \end{figure}
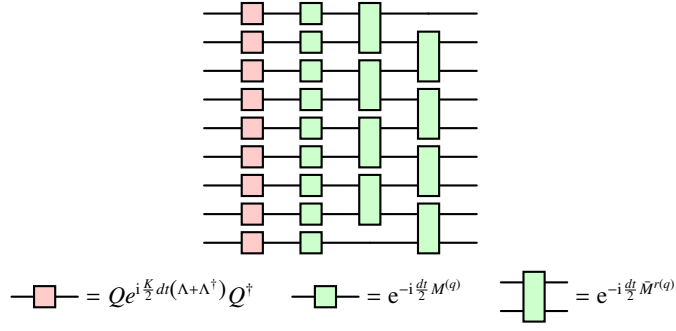

Using the aforementioned operators that can be expressed by using two-qubit gates, we can define the Hamiltonian evolution operator based on the Suzuki-Trotter decomposition:
\begin{equation}
\calU(T)=  \left(e^{-\im dt H_E} e^{-\im dt H_B}\right)^M ,
\end{equation}
\begin{align}
    e^{-\im dt H_E} &=\prod_{x:{\rm odd}} e^{-\im \frac{1}{2}dt \bar{M}^{r(q)}(x,x+1)}\prod_{x:{\rm even}} e^{-\im \frac{1}{2}dt \bar{M}^{r(q)}(x,x+1)}
    \prod_x e^{-\im \frac{1}{2}dt M^{(q)}(x)} ,
\end{align}
\begin{align}
    e^{-\im dt H_B} &=\prod_x Q(x)e^{\im \frac{K}{2}dt \left(\Lambda(x)+\Lambda^\dagger(x)\right)}Q^\dagger(x) .
\end{align}
where $T$, $dt$, and $M=T/dt$ are the time interval, time step, and total time step, respectively. We note that we measure the time scale in units of $c$, and $K$ is rescaled as $K/c\rightarrow K$. We show the circuit diagram using qutrits for illustration purposes in Fig.~\ref{fig:UFcirc}. Quantum simulations are performed using qubits as described in this section.

Although the number of two-qubit gates is largely reduced by using $\bar{M}^{r(q)}$ instead of $M^{r(q)}$, we found that the Hamiltonian evolution is still challenging in present noisy quantum devices. Thus, to study nonequilibrium physics with shorter Trotter steps, we consider Floquet evolution generated by $\calU(T)$ with a large $dt$. Furthermore, we follow Ref.~\cite{Kim2023} and choose $dt=2\pi$, which enables us to reduce the number of the CZ gates used in the $R_{ZZ}$ gates in $e^{-\im dt H_E}$. 
We note also that we use Eqs.~\eqref{eq:casimir}, and~\eqref{eq:casimir_qubit} as $E^2$ operator acting on plaquettes at the ends of a chain i.e., we neglect the difference between Eqs.~\eqref{eq:casimir}, and~\eqref{eq:casimir2}, so as the trick is applicable for all $R_{ZZ}$ gates in $e^{-\im dt H_E}$. 
We study thermalization dynamics with changing the remaining parameter $K$ in experiments described in the next section.

\begin{figure}[t]
    \includegraphics[width=.5\textwidth]
    {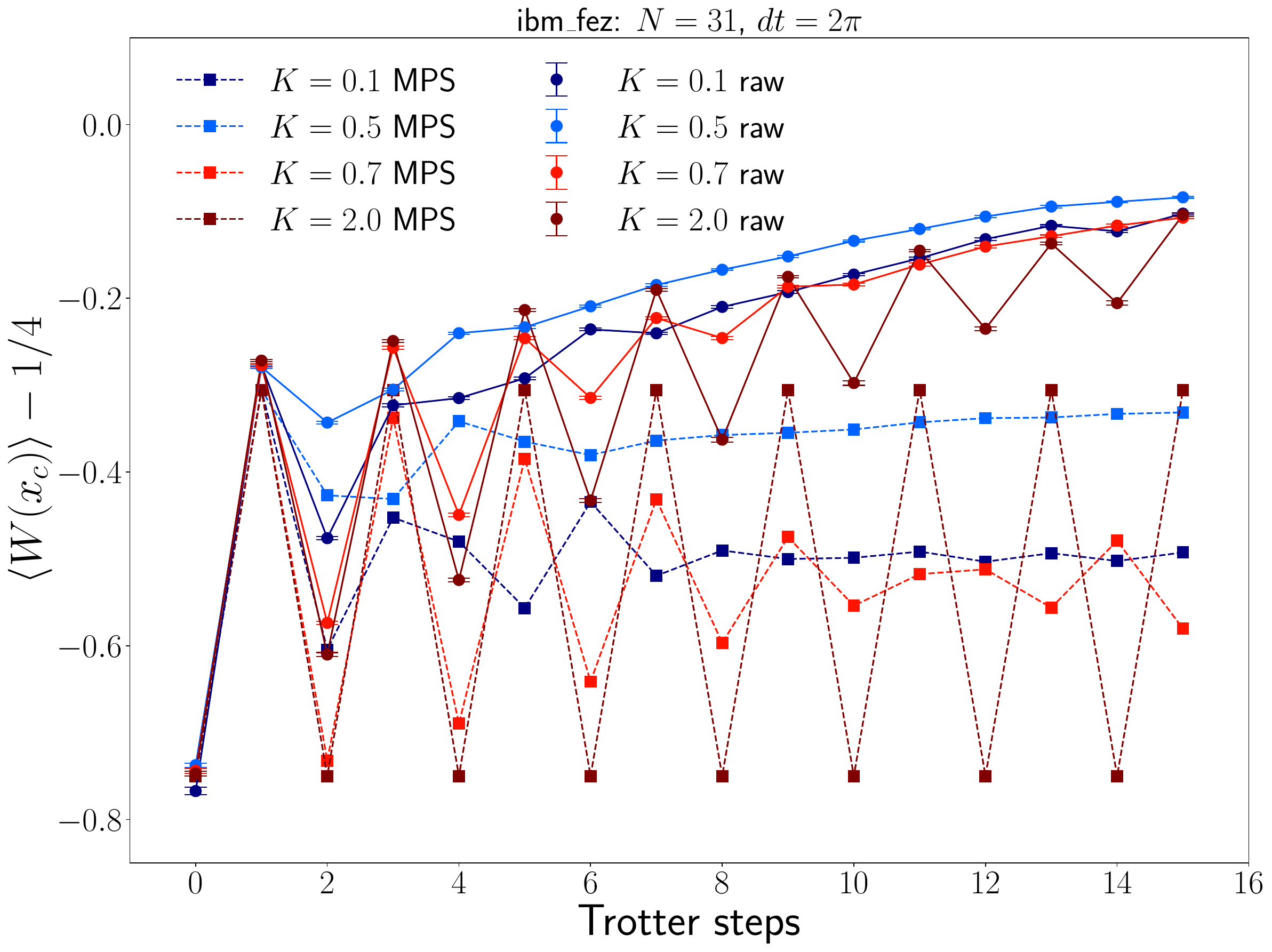}
    \caption{
      \label{fig:raw}
        Quantum simulation results of Trotter-time step dependence of Floquet prethermalization using ibm\_fez. The system size is $N=31$ ($62$ qubits are used on the real device). The coupling constant $K$ is set to $K=0.1$, $0.5$, $0.7$, and $2.0$. ``MPS" shows the results of the classical MPS calculations, while ``raw" shows the results of quantum simulations. Here, the raw data are the results obtained by enabling only TREX and Pauli twirling of two-qubit gates. Lines are just for eye guides. 
    }
\end{figure}

\section{Experiments}
\label{sec:experiments}
We performed experiments using ibm\_fez, IBM's Heron quantum processor consisting of 156 superconducting qubits.
The median CZ, SX, and readout errors were typically $2.96\times10^{-3}$, $2.93\times10^{-4}$, and $1.76\times10^{-2}$, respectively, when we were submitting jobs.
We used Qiskit's estimator~\cite{qiskit_paper} to compute an expectation value of the fundamental Wilson loop operator placed at the center of a $N$-plaquette chain $x_c=(N+1)/2$: $W(x_c)$. We set the number of shots to $100,000$. We enabled Pauli twirling for two-qubit gates and for mitigating readout errors named twirled readout error eXtinction (TREX), while dynamical decoupling was disabled. The number of twirled circuits (\texttt{num\_randomizations} in Qiskit's twirling options) was set to $32$.  We did not use zero noise extrapolation implemented by Qiskit runtime service since the number of two-qubit gates exceeds the system limits in noise amplifications. 

We simulated $31$-plaquette system [i.e., we took $N=31$ in Eqs.~\eqref{eq:magnetic} and~\eqref{eq:electric}], which consists of $62$ qubits in the real device. 
As seen below, we found that the simulations fail due to noises earlier than $15$ Trotter steps even if we perform error mitigation. Then, $N=31$ is large enough to avoid finite-size effects since simulations fail due to noises before the causal cone of $W(x_c)$ reaches the ends of a system. 
The initial state was chosen to be an eigenstate of $W(x)$ that satisfies $W(x)|\Psi(0)\rangle=e^{-i\frac{2\pi}{3}}|\Psi(0)\rangle$ at all $x$, which can be prepared on a quantum circuit as $|\Psi(0)\rangle=\prod_x Q^\dag(x)|0,\cdots\rangle$. We performed simulations with several coupling constants. $K=0.1$, and $0.5$ show prethermal plateaus, while $K=0.7$, and $2.0$ show oscillatory behaviors.

We show the raw experimental data in Fig.~\ref{fig:raw}. We computed Floquet time evolution of the expectation value of $W(x_c)$: $\langle W(x_c)\rangle$ as a function of the number of Trotter steps $M$ with varying $K$. In actual quantum simulations, we change the basis by $Q^\dagger(x_c)$, and measure $0.5\Lambda(x_c)+0.5\Lambda^\dagger(x_c)$ using Qiskit's estimator. To make the depolarising limit clear, we show $\langle W(x_c)\rangle-1/4$ in Fig.~\ref{fig:raw}. $\langle W(x_c)\rangle-1/4$ becomes zero in the depolarising limit (We note that $\Tr[\rho\Lambda(x_c)]=1/4$ with $\rho$ being the maximally mized state $\rho=\bm 1/2^{2N}$).
For comparison, we also performed classical simulations based on matrix product states (MPSs) using ITensor~\cite{itensor}. We simulate the $S=1$ spin chain directly in classical simulations. By changing the cutoff to accuracy in tensor contractions of gates and MPSs, we confirmed that errors due to truncation of singular values are smaller than the size of data points in Fig.~\ref{fig:raw}.

We see that the raw data are qualitatively consistent with classical simulations at very short times. This demonstrates the ability of present superconducting quantum devices to simulate rather complex many-body dynamics (a qutrit chain) than the transverse-field Ising model (a qubit chain). However, as the Trotter time steps increase, the results of quantum computations deviate from those of classical simulations, and go toward the value expected from the depolarizing quantum channel. Thus, error mitigation is inevitable to use quantum processors for simulating a practical quantum many-body problem with much longer Trotter steps. We note that mitigation of readout errors based on TREX is already taken into account in the raw data shown in Fig.~\ref{fig:raw}. 

We perform error mitigation by the following two methods. Let us describe the method and apply it one by one.
We perform error mitigation of physical observables based on the depolarizing noise model, which is a simple noise model widely used in error mitigation. 
In the global depolarizing noise model, the noisy action of a unitary operator $U$ to a $2N$-qubit quantum state $\rho$ $\calD_U(\rho)$ is given by
\begin{align}
    \calD_U(\rho)=f U\rho U^\dag+\frac{1-f}{d}\mathbb{I},
\end{align}
where $d=2^{2N}$ and $1-f$ is the depolarizing error rate.
Then, the expectation value of an observable $O$ is given by
\begin{align}
    \label{eq:expectation_value}
    \langle O\rangle_{\rm noisy}=\Tr[O\calD_U(\rho)]=f\Tr[OU\rho U^\dag]+\frac{1-f}{d}\Tr[O].
\end{align}
Our goal is to correct $\langle O\rangle_{\rm noisy}$ obtained from experiments, and estimate $\langle O\rangle_{\rm ideal}=\Tr[OU\rho U^\dag]$. To this end, we need to estimate $f$ from experiments.

 \begin{figure}[t]
    \includegraphics[width=.5\textwidth]
    {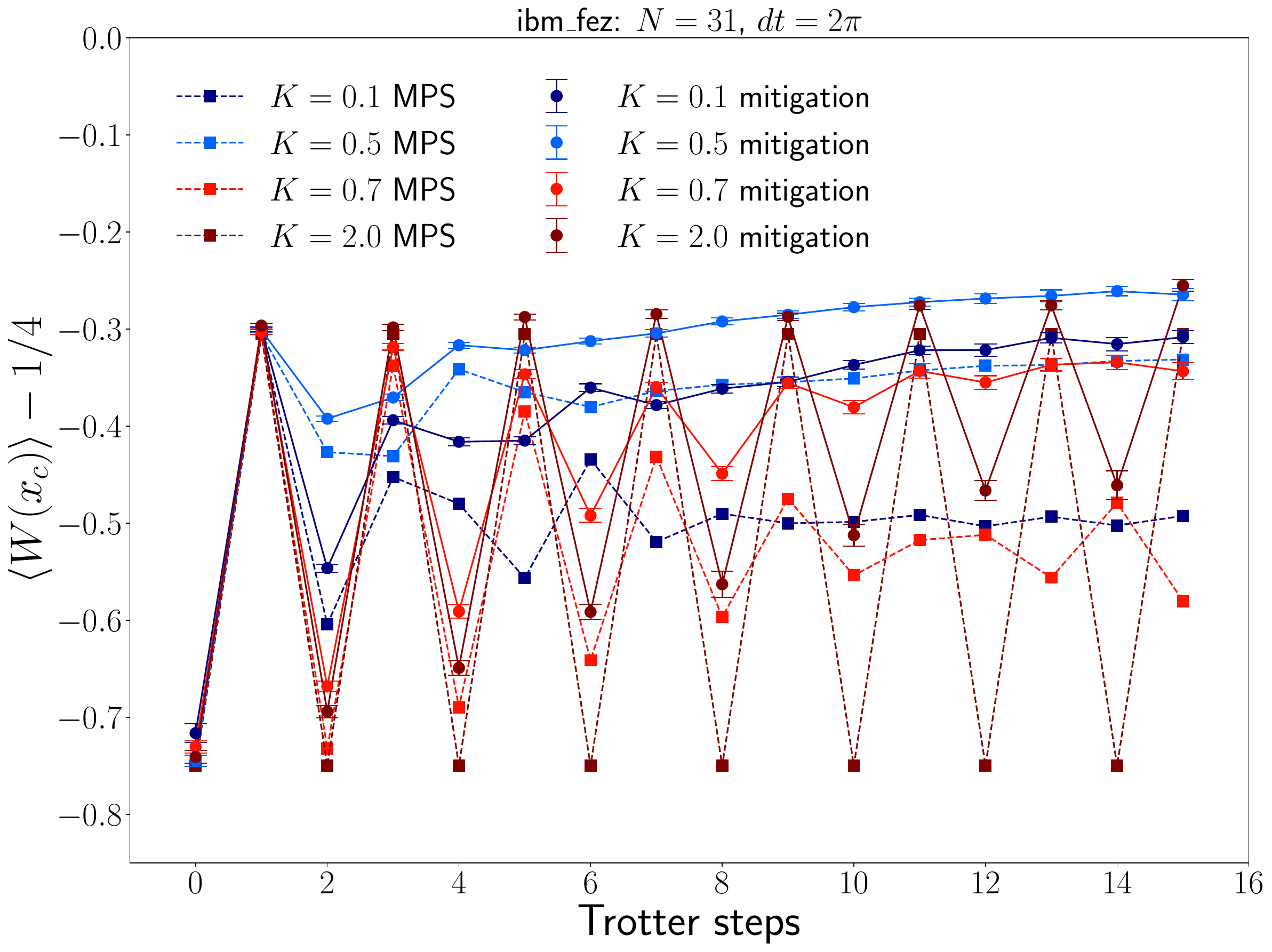}
    \caption{
      \label{fig:gauss}
        Quantum simulation results of Trotter-time step dependence of Floquet prethermalization using ibm\_fez. The system size is $N=31$ ($62$ qubits are used on the real device). The coupling constant $K$ is set to $K=0.1$, $0.5$, $0.7$, and $2.0$. ``MPS" shows the results of the classical MPS calculations, while ``mitigation" shows the results of error mitigation. Here, the rescaling factor is estimated by measuring the projection operator to unused state at the same site. Lines are just for eye guides. 
    }
\end{figure}

First, we estimate $f$ by measuring the projection operator that projects the state to outside of logical qubits. We aim to compute a local expectation value of a 1-qutrit operator at a site $x$.
Since $M_\phi(x)$ is the projection operator to unused state $|11\rangle$, we have
\begin{align}
    \label{eq:projector}
    \langle M_\phi(x)\rangle_{\rm noisy}=\frac{1-f_x}{4}.
\end{align}
Thus, we can estimate the local error rate $f_x$. Using this, we can correct the expectation value of an observable $O(x)$ site by site as
\begin{align}
    \label{eq:expectation_value_mit}
    & \langle O(x)\rangle_{\rm mit} =
    \notag \\
    &\frac{\langle O(x)\rangle_{\rm noisy}}{1-4\langle M_\phi(x)\rangle_{\rm noisy}}-\frac{\langle M_\phi(x)\rangle_{\rm noisy}}{1-4\langle M_\phi(x)\rangle_{\rm noisy}}\Tr[O(x)].
\end{align}
We note that we assume that $f_x$ is independent of observables.
We measured $W(x_c)$ and $P_\phi(x_c)$, which can be done simultaneously, and performed error mitigation based on Eq.~\eqref{eq:expectation_value_mit}. The results are shown in Fig.~\ref{fig:gauss}. We see the agreement between quantum computations and classical simulations becomes better after error mitigation, but there are still quantitative deviations particularly in thermal plateaus. Compared with error mitigation methods presented in previous studies~\cite{Shinjo:2024vci,Hayata:2024smx,Farrell:2024fit}, this error mitigation method fell short of our expectations. This may be due to the fact that $f_x$ is largely different between $W(x_c)$ and $P_\phi(x_c)$. We may need to be careful to choose the mitigation method since we do not know the exact answer in advance in the problems that we want to solve.

\begin{figure}[t]
    \includegraphics[width=.5\textwidth]
    {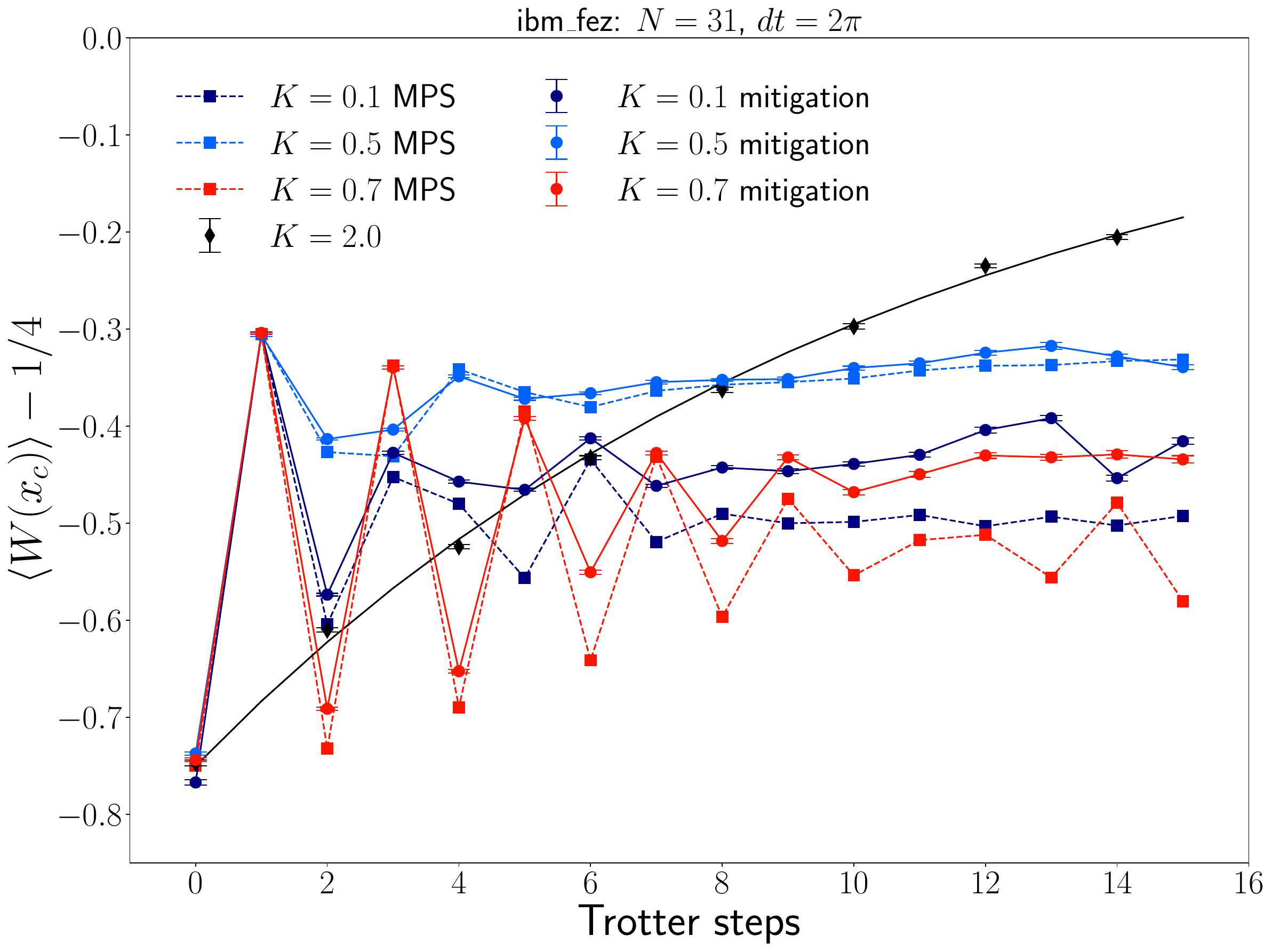}
    \caption{
      \label{fig:odr}
        Quantum simulation results of Trotter-time step dependence of Floquet prethermalization using ibm\_fez. The system size is $N=31$ ($62$ qubits are used on the real device). The coupling constant $K$ is set to $K=0.1$, $0.5$, $0.7$, and $2.0$. ``MPS" shows the results of the classical MPS calculations, while `mitigation" shows the results of error mitigation. Here, the rescaling factor is estimated by running the circuit with the special parameter $K=2.0$. Lines are just for eye guides, while the curve is the result of exponential fitting of $K=2.0$.    
    }
\end{figure}

Second, to estimate a rescaling factor using the same operator that we will measure, we employ the method used in Refs.~\cite{Shinjo:2024vci,Hayata:2024smx}, which is referred to as operator decoherence renormalization in Ref.~\cite{Farrell:2024fit}. 
We simulate the circuit with $K=2.0$, in which $\langle W(x_c)\rangle-\frac{1}{4}$ trivially oscillates like $-3/4\rightarrow -13/36\rightarrow-3/4\rightarrow-13/36\cdots$ if there is no error. 
However, as seen in Fig.~\ref{fig:raw}, the actual data exponentially decreases as the Trotter steps increase.
Following Ref.~\cite{Hayata:2024smx} and using even $M$, we fit the noisy simulations of $K=2.0$ by exponential as
\begin{equation}
\langle W(x_c) \rangle_{\rm raw} =\frac{1}{4}-A e^{-\lambda M}, 
\end{equation}
where $A$ and $\lambda$ are fitting parameters.
Using the fitting results, we perform error mitigation of $K\neq 2.0$ at $M$ as
\begin{equation}
\langle W(x_c) \rangle_{\rm mit} -\frac{1}{4} = \frac{4}{3A} e^{\lambda M}\left(\langle W(x_c) \rangle_{\rm raw}-\frac{1}{4}\right)  .
\end{equation}
The results of error mitigation are shown in Fig.~\ref{fig:odr} (The raw data are shown in Fig.~\ref{fig:raw}). In fact, the second method works remarkably well. Thus, in the situation where zero noise extrapolation is not applicable such as a large circuit, mitigation by operator decoherence renormalization would be the first and possibly best choice as confirmed in previous works~\cite{Shinjo:2024vci,Hayata:2024smx,Farrell:2024fit}. Regarding the quality of mitigation, $K=0.1$ and $0.7$ are almost the same, while $K=0.5$ is unexpectedly good. This may be due to the effective circuit volume may slightly differ in those circuits although the operator is the same. We deepen our understanding of this by computing effective fidelity and quantum volume~\cite{Kechedzhi:2023swt} in the next section.

\begin{figure}[t]
  \centering
  \includegraphics[width=.5\textwidth]
  {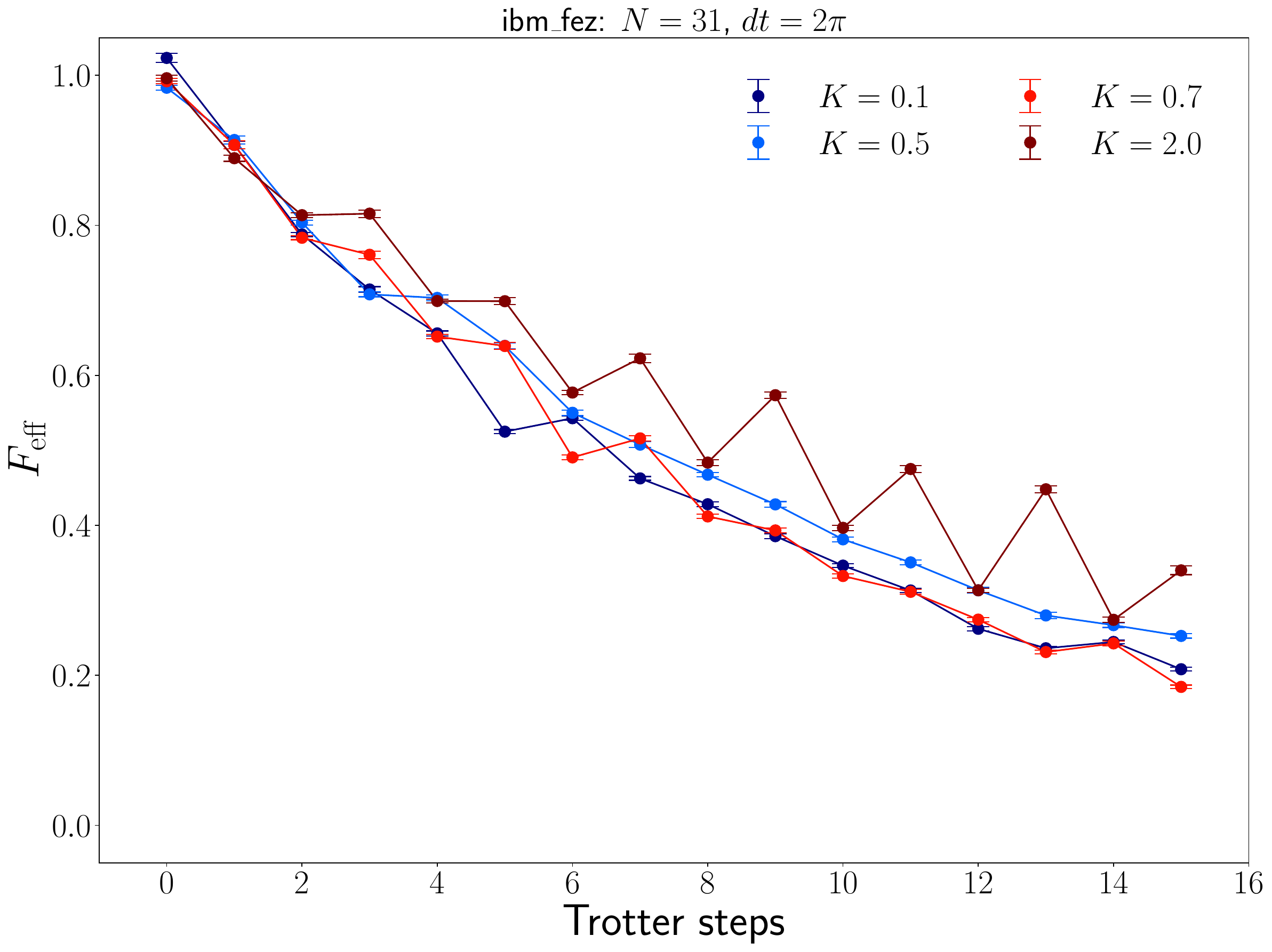}
  \caption{
  \label{fig:fidelity}
    Effective fidelity $F_{\rm eff}$ of quantum circuits simulated using ibm\_fez. For the raw data, see Fig.~\ref{fig:raw}. We estimate $F_{\rm eff}$ by dividing the raw data of quantum simulations by the corresponding ``MPS" results. Lines are just for eye guides. }
    \includegraphics[width=.5\textwidth]
  {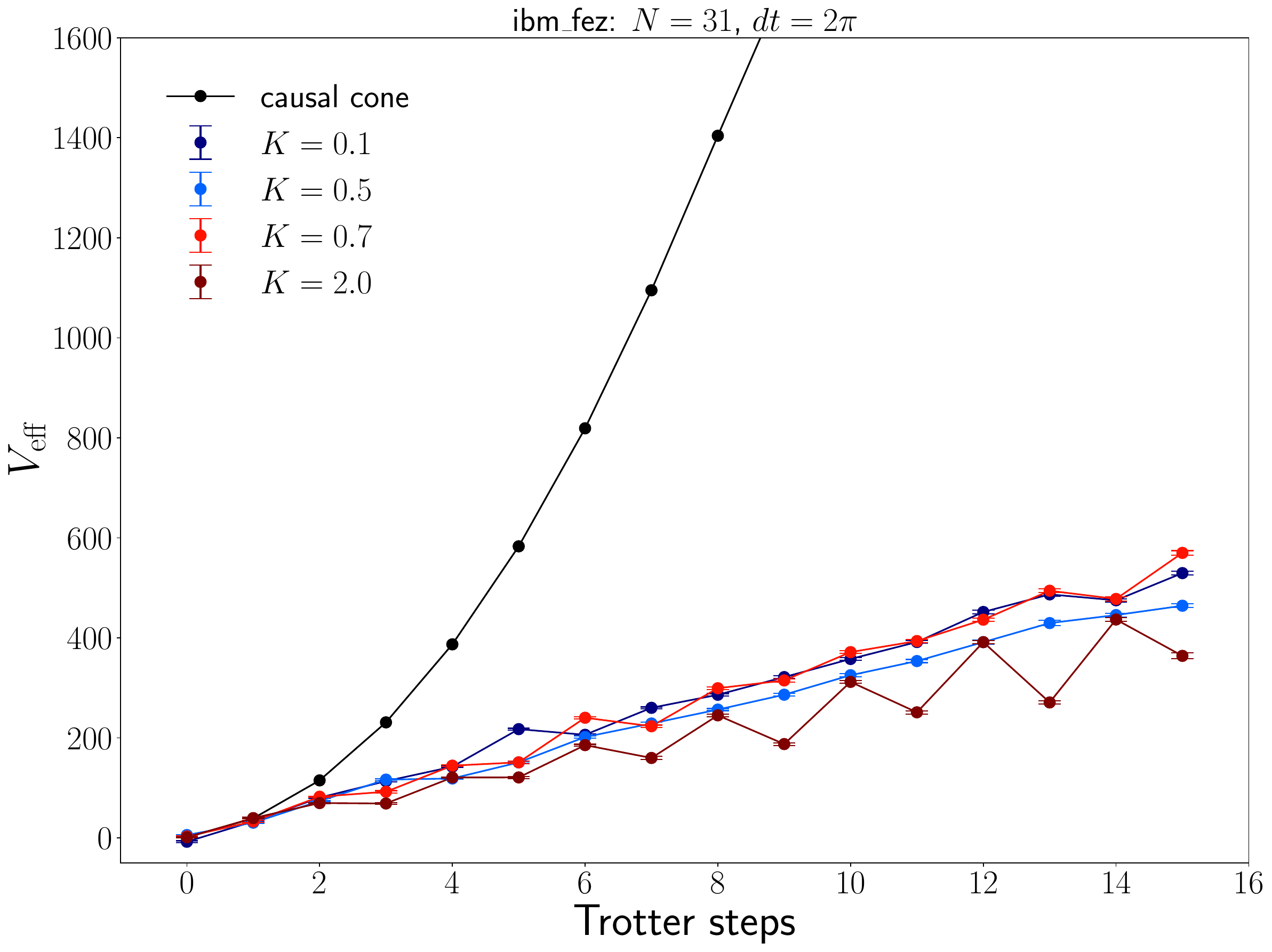}
  \caption{
  \label{fig:volume}
    Effective circuit volume $V_{\rm eff}$ of quantum circuits simulated using ibm\_fez. For the raw data, see Fig.~\ref{fig:raw}. $V_{\rm eff}$ is related to $F_{\rm eff}$ as $F_{\rm eff}=(1-p)^{V_{\rm eff}}$, and we used $p=2.96\times 10^{-3}$, which is the median CZ error of ibm\_fez.   ``causal cone" represents the total number of the CZ gates inside the causal cone of $W(x_c)$. Lines are just for eye guides. }
\end{figure}

\section{Discussion}
\label{sec:conclusion}
We have simulated a truncated SU($3$) lattice Yang-Mills theory on a two-leg ladder geometry using IBM's superconducting 156-qubit device ibm\_fez. To this end,  we have derived the quantum spin representation of the model, which is described as the $S=1$ quantum spin chain, and constructed the explicit mapping of the $S=1$ quantum spin chain to the qubit chain to simulate it on superconducting qubit devices. We have shown that present IBM's quantum device can simulate Floquet thermalization dynamics in the q-deformed SU(3)${}_1$ Yang-Mills theory that consists of more complex quantum circuits than the $\mathbb{Z}_2$ lattice gauge theory~\cite{Hayata:2024smx}. Our work may pave the way to scale up experiments and simulate lattice gauge theories in more physically relevant setups. 

Finally, we discuss the quality of error mitigation in terms of fidelity and entangling gates. To this end, we estimate the error rate $f$ using the raw data of quantum simulations and the MPS results. Following Ref.~\cite{Hayata:2024smx}, we refer to the estimated value as the effective fidelity $F_{\rm eff}$, and the number of entangling gates estimated from $F_{\rm eff}$ as effective circuit volume $V_{\rm eff}$ (see also Ref.~\cite{Kechedzhi:2023swt}). Those are related as $F_{\rm eff}=(1-p)^{V_{\rm eff}}$, where $p$ is typical value of gate errors.
We use the raw data of quantum simulations (error mitigation by TREX is taken into account) as $\langle O\rangle_{\rm noisy}$, the MPS results as $\langle O\rangle_{\rm ideal}$, and estimate $f$ from Eq.~\eqref{eq:expectation_value}. We set $p=2.96\times 10^{-3}$, which is the median CZ error of ibm\_fez when we submitted jobs. 
We show the effective fidelity $F_{\rm eff}$ and circuit volume $V_{\rm eff}$ in Figs.~\ref{fig:fidelity}, and~\ref{fig:volume}, respectively. The quantitative agreement of $F_{\rm eff}$ means that error mitigation by estimating the rescaling factor from the $K=2.0$ circuit works well. In particular, the almost perfect match of decay of $F_{\rm eff}$ between $K=0.5$, and $2.0$ at even Trotter steps explains why error mitigation works remarkably well in this case.
We estimate the number of two-qubit gates involved to expectation values from $V_{\rm eff}$. As seen in Fig.~\ref{fig:volume}, the growth of $V_{\rm eff}$ is much slower than that of the number of two-qubit gates inside causal cones (We used pytket~\cite{tket_paper} to count two-qubit gates inside causal cones). As was observed in the $\mathbb{Z}_2$ lattice gauge theory~\cite{Hayata:2024smx}, this implies that thermalization dynamics in Floquet circuits is  simulatable easily than expected from the counting of the two-qubit gates inside causal cones. Furthermore, from our results and those in Ref.~\cite{Hayata:2024smx}, we conjecture that error mitigation by operator decoherence renormalization may work in ibm\_fez if $V_{\rm eff}$ is less than $400$ two-qubit gates.

\acknowledgments
T.H. thanks Yuta Kikuchi and Kazuhiro Seki for fruitful discussions and critical comments.
The classical simulations were performed using cluster computers at iTHEMS in RIKEN and the quantum simulations were performed using IBM’s superconducting quantum computer on the cloud in IBM Quantum services. This paper is based on results obtained from a project commissioned by the New Energy and Industrial Technology Development Organization (NEDO).
T.~H.~was supported by JSPS KAKENHI Grant No.~24K00630. 
Y.~H.~was partially supported by JSPS KAKENHI Grant Nos.~21H01084 and 24H00975.

\bibliographystyle{apsrev4-2}
\bibliography{scrambling}

\begin{thebibliography}{38}%
\makeatletter
\providecommand \@ifxundefined [1]{%
 \@ifx{#1\undefined}
}%
\providecommand \@ifnum [1]{%
 \ifnum #1\expandafter \@firstoftwo
 \else \expandafter \@secondoftwo
 \fi
}%
\providecommand \@ifx [1]{%
 \ifx #1\expandafter \@firstoftwo
 \else \expandafter \@secondoftwo
 \fi
}%
\providecommand \natexlab [1]{#1}%
\providecommand \enquote  [1]{``#1''}%
\providecommand \bibnamefont  [1]{#1}%
\providecommand \bibfnamefont [1]{#1}%
\providecommand \citenamefont [1]{#1}%
\providecommand \href@noop [0]{\@secondoftwo}%
\providecommand \href [0]{\begingroup \@sanitize@url \@href}%
\providecommand \@href[1]{\@@startlink{#1}\@@href}%
\providecommand \@@href[1]{\endgroup#1\@@endlink}%
\providecommand \@sanitize@url [0]{\catcode `\\12\catcode `\$12\catcode
  `\&12\catcode `\#12\catcode `\^12\catcode `\_12\catcode `\%12\relax}%
\providecommand \@@startlink[1]{}%
\providecommand \@@endlink[0]{}%
\providecommand \url  [0]{\begingroup\@sanitize@url \@url }%
\providecommand \@url [1]{\endgroup\@href {#1}{\urlprefix }}%
\providecommand \urlprefix  [0]{URL }%
\providecommand \Eprint [0]{\href }%
\providecommand \doibase [0]{https://doi.org/}%
\providecommand \selectlanguage [0]{\@gobble}%
\providecommand \bibinfo  [0]{\@secondoftwo}%
\providecommand \bibfield  [0]{\@secondoftwo}%
\providecommand \translation [1]{[#1]}%
\providecommand \BibitemOpen [0]{}%
\providecommand \bibitemStop [0]{}%
\providecommand \bibitemNoStop [0]{.\EOS\space}%
\providecommand \EOS [0]{\spacefactor3000\relax}%
\providecommand \BibitemShut  [1]{\csname bibitem#1\endcsname}%
\let\auto@bib@innerbib\@empty
\bibitem [{\citenamefont {Klco}\ \emph {et~al.}(2018)\citenamefont {Klco},
  \citenamefont {Dumitrescu}, \citenamefont {McCaskey}, \citenamefont {Morris},
  \citenamefont {Pooser}, \citenamefont {Sanz}, \citenamefont {Solano},
  \citenamefont {Lougovski},\ and\ \citenamefont {Savage}}]{Klco:2018kyo}%
  \BibitemOpen
  \bibfield  {author} {\bibinfo {author} {\bibfnamefont {N.}~\bibnamefont
  {Klco}}, \bibinfo {author} {\bibfnamefont {E.~F.}\ \bibnamefont
  {Dumitrescu}}, \bibinfo {author} {\bibfnamefont {A.~J.}\ \bibnamefont
  {McCaskey}}, \bibinfo {author} {\bibfnamefont {T.~D.}\ \bibnamefont
  {Morris}}, \bibinfo {author} {\bibfnamefont {R.~C.}\ \bibnamefont {Pooser}},
  \bibinfo {author} {\bibfnamefont {M.}~\bibnamefont {Sanz}}, \bibinfo {author}
  {\bibfnamefont {E.}~\bibnamefont {Solano}}, \bibinfo {author} {\bibfnamefont
  {P.}~\bibnamefont {Lougovski}},\ and\ \bibinfo {author} {\bibfnamefont
  {M.~J.}\ \bibnamefont {Savage}},\ }\href
  {https://doi.org/10.1103/PhysRevA.98.032331} {\bibfield  {journal} {\bibinfo
  {journal} {Phys. Rev. A}\ }\textbf {\bibinfo {volume} {98}},\ \bibinfo
  {pages} {032331} (\bibinfo {year} {2018})},\ \Eprint
  {https://arxiv.org/abs/1803.03326} {arXiv:1803.03326 [quant-ph]} \BibitemShut
  {NoStop}%
\bibitem [{\citenamefont {Klco}\ \emph {et~al.}(2020)\citenamefont {Klco},
  \citenamefont {Stryker},\ and\ \citenamefont {Savage}}]{Klco:2019evd}%
  \BibitemOpen
  \bibfield  {author} {\bibinfo {author} {\bibfnamefont {N.}~\bibnamefont
  {Klco}}, \bibinfo {author} {\bibfnamefont {J.~R.}\ \bibnamefont {Stryker}},\
  and\ \bibinfo {author} {\bibfnamefont {M.~J.}\ \bibnamefont {Savage}},\
  }\href {https://doi.org/10.1103/PhysRevD.101.074512} {\bibfield  {journal}
  {\bibinfo  {journal} {Phys. Rev. D}\ }\textbf {\bibinfo {volume} {101}},\
  \bibinfo {pages} {074512} (\bibinfo {year} {2020})},\ \Eprint
  {https://arxiv.org/abs/1908.06935} {arXiv:1908.06935 [quant-ph]} \BibitemShut
  {NoStop}%
\bibitem [{\citenamefont {Ciavarella}\ \emph {et~al.}(2021)\citenamefont
  {Ciavarella}, \citenamefont {Klco},\ and\ \citenamefont
  {Savage}}]{Ciavarella:2021nmj}%
  \BibitemOpen
  \bibfield  {author} {\bibinfo {author} {\bibfnamefont {A.}~\bibnamefont
  {Ciavarella}}, \bibinfo {author} {\bibfnamefont {N.}~\bibnamefont {Klco}},\
  and\ \bibinfo {author} {\bibfnamefont {M.~J.}\ \bibnamefont {Savage}},\
  }\href {https://doi.org/10.1103/PhysRevD.103.094501} {\bibfield  {journal}
  {\bibinfo  {journal} {Phys. Rev. D}\ }\textbf {\bibinfo {volume} {103}},\
  \bibinfo {pages} {094501} (\bibinfo {year} {2021})},\ \Eprint
  {https://arxiv.org/abs/2101.10227} {arXiv:2101.10227 [quant-ph]} \BibitemShut
  {NoStop}%
\bibitem [{\citenamefont {Atas}\ \emph {et~al.}(2021)\citenamefont {Atas},
  \citenamefont {Zhang}, \citenamefont {Lewis}, \citenamefont {Jahanpour},
  \citenamefont {Haase},\ and\ \citenamefont {Muschik}}]{Atas:2021ext}%
  \BibitemOpen
  \bibfield  {author} {\bibinfo {author} {\bibfnamefont {Y.~Y.}\ \bibnamefont
  {Atas}}, \bibinfo {author} {\bibfnamefont {J.}~\bibnamefont {Zhang}},
  \bibinfo {author} {\bibfnamefont {R.}~\bibnamefont {Lewis}}, \bibinfo
  {author} {\bibfnamefont {A.}~\bibnamefont {Jahanpour}}, \bibinfo {author}
  {\bibfnamefont {J.~F.}\ \bibnamefont {Haase}},\ and\ \bibinfo {author}
  {\bibfnamefont {C.~A.}\ \bibnamefont {Muschik}},\ }\href
  {https://doi.org/10.1038/s41467-021-26825-4} {\bibfield  {journal} {\bibinfo
  {journal} {Nature Commun.}\ }\textbf {\bibinfo {volume} {12}},\ \bibinfo
  {pages} {6499} (\bibinfo {year} {2021})},\ \Eprint
  {https://arxiv.org/abs/2102.08920} {arXiv:2102.08920 [quant-ph]} \BibitemShut
  {NoStop}%
\bibitem [{\citenamefont {de~Jong}\ \emph {et~al.}(2022)\citenamefont
  {de~Jong}, \citenamefont {Lee}, \citenamefont {Mulligan}, \citenamefont
  {P\l{}osko\'n}, \citenamefont {Ringer},\ and\ \citenamefont
  {Yao}}]{deJong:2021wsd}%
  \BibitemOpen
  \bibfield  {author} {\bibinfo {author} {\bibfnamefont {W.~A.}\ \bibnamefont
  {de~Jong}}, \bibinfo {author} {\bibfnamefont {K.}~\bibnamefont {Lee}},
  \bibinfo {author} {\bibfnamefont {J.}~\bibnamefont {Mulligan}}, \bibinfo
  {author} {\bibfnamefont {M.}~\bibnamefont {P\l{}osko\'n}}, \bibinfo {author}
  {\bibfnamefont {F.}~\bibnamefont {Ringer}},\ and\ \bibinfo {author}
  {\bibfnamefont {X.}~\bibnamefont {Yao}},\ }\href
  {https://doi.org/10.1103/PhysRevD.106.054508} {\bibfield  {journal} {\bibinfo
   {journal} {Phys. Rev. D}\ }\textbf {\bibinfo {volume} {106}},\ \bibinfo
  {pages} {054508} (\bibinfo {year} {2022})},\ \Eprint
  {https://arxiv.org/abs/2106.08394} {arXiv:2106.08394 [quant-ph]} \BibitemShut
  {NoStop}%
\bibitem [{\citenamefont {Ciavarella}\ and\ \citenamefont
  {Chernyshev}(2022)}]{Ciavarella:2021lel}%
  \BibitemOpen
  \bibfield  {author} {\bibinfo {author} {\bibfnamefont {A.~N.}\ \bibnamefont
  {Ciavarella}}\ and\ \bibinfo {author} {\bibfnamefont {I.~A.}\ \bibnamefont
  {Chernyshev}},\ }\href {https://doi.org/10.1103/PhysRevD.105.074504}
  {\bibfield  {journal} {\bibinfo  {journal} {Phys. Rev. D}\ }\textbf {\bibinfo
  {volume} {105}},\ \bibinfo {pages} {074504} (\bibinfo {year} {2022})},\
  \Eprint {https://arxiv.org/abs/2112.09083} {arXiv:2112.09083 [quant-ph]}
  \BibitemShut {NoStop}%
\bibitem [{\citenamefont {Nguyen}\ \emph {et~al.}(2022)\citenamefont {Nguyen},
  \citenamefont {Tran}, \citenamefont {Zhu}, \citenamefont {Green},
  \citenamefont {Alderete}, \citenamefont {Davoudi},\ and\ \citenamefont
  {Linke}}]{Nguyen:2021hyk}%
  \BibitemOpen
  \bibfield  {author} {\bibinfo {author} {\bibfnamefont {N.~H.}\ \bibnamefont
  {Nguyen}}, \bibinfo {author} {\bibfnamefont {M.~C.}\ \bibnamefont {Tran}},
  \bibinfo {author} {\bibfnamefont {Y.}~\bibnamefont {Zhu}}, \bibinfo {author}
  {\bibfnamefont {A.~M.}\ \bibnamefont {Green}}, \bibinfo {author}
  {\bibfnamefont {C.~H.}\ \bibnamefont {Alderete}}, \bibinfo {author}
  {\bibfnamefont {Z.}~\bibnamefont {Davoudi}},\ and\ \bibinfo {author}
  {\bibfnamefont {N.~M.}\ \bibnamefont {Linke}},\ }\href
  {https://doi.org/10.1103/PRXQuantum.3.020324} {\bibfield  {journal} {\bibinfo
   {journal} {PRX Quantum}\ }\textbf {\bibinfo {volume} {3}},\ \bibinfo {pages}
  {020324} (\bibinfo {year} {2022})},\ \Eprint
  {https://arxiv.org/abs/2112.14262} {arXiv:2112.14262 [quant-ph]} \BibitemShut
  {NoStop}%
\bibitem [{\citenamefont {Atas}\ \emph {et~al.}(2023)\citenamefont {Atas},
  \citenamefont {Haase}, \citenamefont {Zhang}, \citenamefont {Wei},
  \citenamefont {Pfaendler}, \citenamefont {Lewis},\ and\ \citenamefont
  {Muschik}}]{Atas:2022dqm}%
  \BibitemOpen
  \bibfield  {author} {\bibinfo {author} {\bibfnamefont {Y.~Y.}\ \bibnamefont
  {Atas}}, \bibinfo {author} {\bibfnamefont {J.~F.}\ \bibnamefont {Haase}},
  \bibinfo {author} {\bibfnamefont {J.}~\bibnamefont {Zhang}}, \bibinfo
  {author} {\bibfnamefont {V.}~\bibnamefont {Wei}}, \bibinfo {author}
  {\bibfnamefont {S.~M.~L.}\ \bibnamefont {Pfaendler}}, \bibinfo {author}
  {\bibfnamefont {R.}~\bibnamefont {Lewis}},\ and\ \bibinfo {author}
  {\bibfnamefont {C.~A.}\ \bibnamefont {Muschik}},\ }\href
  {https://doi.org/10.1103/PhysRevResearch.5.033184} {\bibfield  {journal}
  {\bibinfo  {journal} {Phys. Rev. Res.}\ }\textbf {\bibinfo {volume} {5}},\
  \bibinfo {pages} {033184} (\bibinfo {year} {2023})},\ \Eprint
  {https://arxiv.org/abs/2207.03473} {arXiv:2207.03473 [quant-ph]} \BibitemShut
  {NoStop}%
\bibitem [{\citenamefont {A~Rahman}\ \emph {et~al.}(2022)\citenamefont
  {A~Rahman}, \citenamefont {Lewis}, \citenamefont {Mendicelli},\ and\
  \citenamefont {Powell}}]{ARahman:2022tkr}%
  \BibitemOpen
  \bibfield  {author} {\bibinfo {author} {\bibfnamefont {S.}~\bibnamefont
  {A~Rahman}}, \bibinfo {author} {\bibfnamefont {R.}~\bibnamefont {Lewis}},
  \bibinfo {author} {\bibfnamefont {E.}~\bibnamefont {Mendicelli}},\ and\
  \bibinfo {author} {\bibfnamefont {S.}~\bibnamefont {Powell}},\ }\href
  {https://doi.org/10.1103/PhysRevD.106.074502} {\bibfield  {journal} {\bibinfo
   {journal} {Phys. Rev. D}\ }\textbf {\bibinfo {volume} {106}},\ \bibinfo
  {pages} {074502} (\bibinfo {year} {2022})},\ \Eprint
  {https://arxiv.org/abs/2205.09247} {arXiv:2205.09247 [hep-lat]} \BibitemShut
  {NoStop}%
\bibitem [{\citenamefont {Charles}\ \emph {et~al.}(2024)\citenamefont
  {Charles}, \citenamefont {Gustafson}, \citenamefont {Hardt}, \citenamefont
  {Herren}, \citenamefont {Hogan}, \citenamefont {Lamm}, \citenamefont
  {Starecheski}, \citenamefont {Van~de Water},\ and\ \citenamefont
  {Wagman}}]{Charles:2023zbl}%
  \BibitemOpen
  \bibfield  {author} {\bibinfo {author} {\bibfnamefont {C.}~\bibnamefont
  {Charles}}, \bibinfo {author} {\bibfnamefont {E.~J.}\ \bibnamefont
  {Gustafson}}, \bibinfo {author} {\bibfnamefont {E.}~\bibnamefont {Hardt}},
  \bibinfo {author} {\bibfnamefont {F.}~\bibnamefont {Herren}}, \bibinfo
  {author} {\bibfnamefont {N.}~\bibnamefont {Hogan}}, \bibinfo {author}
  {\bibfnamefont {H.}~\bibnamefont {Lamm}}, \bibinfo {author} {\bibfnamefont
  {S.}~\bibnamefont {Starecheski}}, \bibinfo {author} {\bibfnamefont {R.~S.}\
  \bibnamefont {Van~de Water}},\ and\ \bibinfo {author} {\bibfnamefont {M.~L.}\
  \bibnamefont {Wagman}},\ }\href {https://doi.org/10.1103/PhysRevE.109.015307}
  {\bibfield  {journal} {\bibinfo  {journal} {Phys. Rev. E}\ }\textbf {\bibinfo
  {volume} {109}},\ \bibinfo {pages} {015307} (\bibinfo {year} {2024})},\
  \Eprint {https://arxiv.org/abs/2305.02361} {arXiv:2305.02361 [hep-lat]}
  \BibitemShut {NoStop}%
\bibitem [{\citenamefont {Farrell}\ \emph
  {et~al.}(2024{\natexlab{a}})\citenamefont {Farrell}, \citenamefont {Illa},
  \citenamefont {Ciavarella},\ and\ \citenamefont {Savage}}]{Farrell:2023fgd}%
  \BibitemOpen
  \bibfield  {author} {\bibinfo {author} {\bibfnamefont {R.~C.}\ \bibnamefont
  {Farrell}}, \bibinfo {author} {\bibfnamefont {M.}~\bibnamefont {Illa}},
  \bibinfo {author} {\bibfnamefont {A.~N.}\ \bibnamefont {Ciavarella}},\ and\
  \bibinfo {author} {\bibfnamefont {M.~J.}\ \bibnamefont {Savage}},\ }\href
  {https://doi.org/10.1103/PRXQuantum.5.020315} {\bibfield  {journal} {\bibinfo
   {journal} {PRX Quantum}\ }\textbf {\bibinfo {volume} {5}},\ \bibinfo {pages}
  {020315} (\bibinfo {year} {2024}{\natexlab{a}})},\ \Eprint
  {https://arxiv.org/abs/2308.04481} {arXiv:2308.04481 [quant-ph]} \BibitemShut
  {NoStop}%
\bibitem [{\citenamefont {Schuster}\ \emph {et~al.}(2024)\citenamefont
  {Schuster}, \citenamefont {K\"uhn}, \citenamefont {Funcke}, \citenamefont
  {Hartung}, \citenamefont {Pleinert}, \citenamefont {von Zanthier},\ and\
  \citenamefont {Jansen}}]{Schuster:2023klj}%
  \BibitemOpen
  \bibfield  {author} {\bibinfo {author} {\bibfnamefont {S.}~\bibnamefont
  {Schuster}}, \bibinfo {author} {\bibfnamefont {S.}~\bibnamefont {K\"uhn}},
  \bibinfo {author} {\bibfnamefont {L.}~\bibnamefont {Funcke}}, \bibinfo
  {author} {\bibfnamefont {T.}~\bibnamefont {Hartung}}, \bibinfo {author}
  {\bibfnamefont {M.-O.}\ \bibnamefont {Pleinert}}, \bibinfo {author}
  {\bibfnamefont {J.}~\bibnamefont {von Zanthier}},\ and\ \bibinfo {author}
  {\bibfnamefont {K.}~\bibnamefont {Jansen}},\ }\href
  {https://doi.org/10.1103/PhysRevD.109.114508} {\bibfield  {journal} {\bibinfo
   {journal} {Phys. Rev. D}\ }\textbf {\bibinfo {volume} {109}},\ \bibinfo
  {pages} {114508} (\bibinfo {year} {2024})},\ \Eprint
  {https://arxiv.org/abs/2311.14825} {arXiv:2311.14825 [hep-lat]} \BibitemShut
  {NoStop}%
\bibitem [{\citenamefont {Angelides}\ \emph {et~al.}(2023)\citenamefont
  {Angelides}, \citenamefont {Naredi}, \citenamefont {Crippa}, \citenamefont
  {Jansen}, \citenamefont {K\"uhn}, \citenamefont {Tavernelli},\ and\
  \citenamefont {Wang}}]{Angelides:2023noe}%
  \BibitemOpen
  \bibfield  {author} {\bibinfo {author} {\bibfnamefont {T.}~\bibnamefont
  {Angelides}}, \bibinfo {author} {\bibfnamefont {P.}~\bibnamefont {Naredi}},
  \bibinfo {author} {\bibfnamefont {A.}~\bibnamefont {Crippa}}, \bibinfo
  {author} {\bibfnamefont {K.}~\bibnamefont {Jansen}}, \bibinfo {author}
  {\bibfnamefont {S.}~\bibnamefont {K\"uhn}}, \bibinfo {author} {\bibfnamefont
  {I.}~\bibnamefont {Tavernelli}},\ and\ \bibinfo {author} {\bibfnamefont
  {D.~S.}\ \bibnamefont {Wang}},\ }\href@noop {} {\bibfield  {journal}
  {\bibinfo  {journal} {arXiv preprint}\ } (\bibinfo {year} {2023})},\ \Eprint
  {https://arxiv.org/abs/2312.12831} {arXiv:2312.12831 [hep-lat]} \BibitemShut
  {NoStop}%
\bibitem [{\citenamefont {Farrell}\ \emph
  {et~al.}(2024{\natexlab{b}})\citenamefont {Farrell}, \citenamefont {Illa},
  \citenamefont {Ciavarella},\ and\ \citenamefont {Savage}}]{Farrell:2024fit}%
  \BibitemOpen
  \bibfield  {author} {\bibinfo {author} {\bibfnamefont {R.~C.}\ \bibnamefont
  {Farrell}}, \bibinfo {author} {\bibfnamefont {M.}~\bibnamefont {Illa}},
  \bibinfo {author} {\bibfnamefont {A.~N.}\ \bibnamefont {Ciavarella}},\ and\
  \bibinfo {author} {\bibfnamefont {M.~J.}\ \bibnamefont {Savage}},\ }\href
  {https://doi.org/10.1103/PhysRevD.109.114510} {\bibfield  {journal} {\bibinfo
   {journal} {Phys. Rev. D}\ }\textbf {\bibinfo {volume} {109}},\ \bibinfo
  {pages} {114510} (\bibinfo {year} {2024}{\natexlab{b}})},\ \Eprint
  {https://arxiv.org/abs/2401.08044} {arXiv:2401.08044 [quant-ph]} \BibitemShut
  {NoStop}%
\bibitem [{\citenamefont {Ciavarella}\ and\ \citenamefont
  {Bauer}(2024)}]{Ciavarella:2024fzw}%
  \BibitemOpen
  \bibfield  {author} {\bibinfo {author} {\bibfnamefont {A.~N.}\ \bibnamefont
  {Ciavarella}}\ and\ \bibinfo {author} {\bibfnamefont {C.~W.}\ \bibnamefont
  {Bauer}},\ }\href {https://doi.org/10.1103/PhysRevLett.133.111901} {\bibfield
   {journal} {\bibinfo  {journal} {Phys. Rev. Lett.}\ }\textbf {\bibinfo
  {volume} {133}},\ \bibinfo {pages} {111901} (\bibinfo {year} {2024})},\
  \Eprint {https://arxiv.org/abs/2402.10265} {arXiv:2402.10265 [hep-ph]}
  \BibitemShut {NoStop}%
\bibitem [{\citenamefont {Hayata}\ \emph {et~al.}(2024)\citenamefont {Hayata},
  \citenamefont {Seki},\ and\ \citenamefont {Yamamoto}}]{Hayata:2024smx}%
  \BibitemOpen
  \bibfield  {author} {\bibinfo {author} {\bibfnamefont {T.}~\bibnamefont
  {Hayata}}, \bibinfo {author} {\bibfnamefont {K.}~\bibnamefont {Seki}},\ and\
  \bibinfo {author} {\bibfnamefont {A.}~\bibnamefont {Yamamoto}},\ }\href@noop
  {} {\bibinfo {title} {{Floquet prethermalization of ${\bf Z}_2$ lattice gauge
  theory on superconducting qubits}}} (\bibinfo {year} {2024}),\ \Eprint
  {https://arxiv.org/abs/2408.10079} {arXiv:2408.10079 [hep-lat]} \BibitemShut
  {NoStop}%
\bibitem [{\citenamefont {Cochran}\ \emph {et~al.}(2024)\citenamefont {Cochran}
  \emph {et~al.}}]{Cochran:2024rwe}%
  \BibitemOpen
  \bibfield  {author} {\bibinfo {author} {\bibfnamefont {T.~A.}\ \bibnamefont
  {Cochran}} \emph {et~al.},\ }\href@noop {} {\bibinfo {title} {{Visualizing
  Dynamics of Charges and Strings in (2+1)D Lattice Gauge Theories}}} (\bibinfo
  {year} {2024}),\ \Eprint {https://arxiv.org/abs/2409.17142} {arXiv:2409.17142
  [quant-ph]} \BibitemShut {NoStop}%
\bibitem [{\citenamefont {Eckstein}\ \emph {et~al.}(2023)\citenamefont
  {Eckstein}, \citenamefont {Mansuroglu}, \citenamefont {Czarnik},
  \citenamefont {Zhu}, \citenamefont {Hartmann}, \citenamefont {Cincio},
  \citenamefont {Sornborger},\ and\ \citenamefont {Holmes}}]{Eckstein:2023sjk}%
  \BibitemOpen
  \bibfield  {author} {\bibinfo {author} {\bibfnamefont {T.}~\bibnamefont
  {Eckstein}}, \bibinfo {author} {\bibfnamefont {R.}~\bibnamefont
  {Mansuroglu}}, \bibinfo {author} {\bibfnamefont {P.}~\bibnamefont {Czarnik}},
  \bibinfo {author} {\bibfnamefont {J.-X.}\ \bibnamefont {Zhu}}, \bibinfo
  {author} {\bibfnamefont {M.~J.}\ \bibnamefont {Hartmann}}, \bibinfo {author}
  {\bibfnamefont {L.}~\bibnamefont {Cincio}}, \bibinfo {author} {\bibfnamefont
  {A.~T.}\ \bibnamefont {Sornborger}},\ and\ \bibinfo {author} {\bibfnamefont
  {Z.}~\bibnamefont {Holmes}},\ }\href@noop {} {\bibfield  {journal} {\bibinfo
  {journal} {arXiv preprint}\ } (\bibinfo {year} {2023})},\ \Eprint
  {https://arxiv.org/abs/2303.02209} {arXiv:2303.02209 [quant-ph]} \BibitemShut
  {NoStop}%
\bibitem [{\citenamefont {Yang}\ \emph {et~al.}(2023)\citenamefont {Yang},
  \citenamefont {Christianen}, \citenamefont {Coll-Vinent}, \citenamefont
  {Smelyanskiy}, \citenamefont {Ba\~nuls}, \citenamefont {O'Brien},
  \citenamefont {Wild},\ and\ \citenamefont {Cirac}}]{Yang:2023nak}%
  \BibitemOpen
  \bibfield  {author} {\bibinfo {author} {\bibfnamefont {Y.}~\bibnamefont
  {Yang}}, \bibinfo {author} {\bibfnamefont {A.}~\bibnamefont {Christianen}},
  \bibinfo {author} {\bibfnamefont {S.}~\bibnamefont {Coll-Vinent}}, \bibinfo
  {author} {\bibfnamefont {V.}~\bibnamefont {Smelyanskiy}}, \bibinfo {author}
  {\bibfnamefont {M.~C.}\ \bibnamefont {Ba\~nuls}}, \bibinfo {author}
  {\bibfnamefont {T.~E.}\ \bibnamefont {O'Brien}}, \bibinfo {author}
  {\bibfnamefont {D.~S.}\ \bibnamefont {Wild}},\ and\ \bibinfo {author}
  {\bibfnamefont {J.~I.}\ \bibnamefont {Cirac}},\ }\href
  {https://doi.org/10.1103/PRXQuantum.4.030320} {\bibfield  {journal} {\bibinfo
   {journal} {PRX Quantum}\ }\textbf {\bibinfo {volume} {4}},\ \bibinfo {pages}
  {030320} (\bibinfo {year} {2023})},\ \Eprint
  {https://arxiv.org/abs/2303.08461} {arXiv:2303.08461 [quant-ph]} \BibitemShut
  {NoStop}%
\bibitem [{\citenamefont {Shinjo}\ \emph {et~al.}(2024)\citenamefont {Shinjo},
  \citenamefont {Seki}, \citenamefont {Shirakawa}, \citenamefont {Sun},\ and\
  \citenamefont {Yunoki}}]{Shinjo:2024vci}%
  \BibitemOpen
  \bibfield  {author} {\bibinfo {author} {\bibfnamefont {K.}~\bibnamefont
  {Shinjo}}, \bibinfo {author} {\bibfnamefont {K.}~\bibnamefont {Seki}},
  \bibinfo {author} {\bibfnamefont {T.}~\bibnamefont {Shirakawa}}, \bibinfo
  {author} {\bibfnamefont {R.-Y.}\ \bibnamefont {Sun}},\ and\ \bibinfo {author}
  {\bibfnamefont {S.}~\bibnamefont {Yunoki}},\ }\href@noop {} {\bibfield
  {journal} {\bibinfo  {journal} {arXiv preprint}\ } (\bibinfo {year}
  {2024})},\ \Eprint {https://arxiv.org/abs/2403.16718} {arXiv:2403.16718
  [quant-ph]} \BibitemShut {NoStop}%
\bibitem [{\citenamefont {Seki}\ \emph {et~al.}(2024)\citenamefont {Seki},
  \citenamefont {Kikuchi}, \citenamefont {Hayata},\ and\ \citenamefont
  {Yunoki}}]{Seki:2024rfx}%
  \BibitemOpen
  \bibfield  {author} {\bibinfo {author} {\bibfnamefont {K.}~\bibnamefont
  {Seki}}, \bibinfo {author} {\bibfnamefont {Y.}~\bibnamefont {Kikuchi}},
  \bibinfo {author} {\bibfnamefont {T.}~\bibnamefont {Hayata}},\ and\ \bibinfo
  {author} {\bibfnamefont {S.}~\bibnamefont {Yunoki}},\ }\href@noop {}
  {\bibinfo {title} {{Simulating Floquet scrambling circuits on trapped-ion
  quantum computers}}} (\bibinfo {year} {2024}),\ \Eprint
  {https://arxiv.org/abs/2405.07613} {arXiv:2405.07613 [quant-ph]} \BibitemShut
  {NoStop}%
\bibitem [{\citenamefont {Lazarides}\ \emph {et~al.}(2014)\citenamefont
  {Lazarides}, \citenamefont {Das},\ and\ \citenamefont
  {Moessner}}]{Lazarides2014}%
  \BibitemOpen
  \bibfield  {author} {\bibinfo {author} {\bibfnamefont {A.}~\bibnamefont
  {Lazarides}}, \bibinfo {author} {\bibfnamefont {A.}~\bibnamefont {Das}},\
  and\ \bibinfo {author} {\bibfnamefont {R.}~\bibnamefont {Moessner}},\ }\href
  {https://doi.org/10.1103/PhysRevE.90.012110} {\bibfield  {journal} {\bibinfo
  {journal} {Phys. Rev. E}\ }\textbf {\bibinfo {volume} {90}},\ \bibinfo
  {pages} {012110} (\bibinfo {year} {2014})}\BibitemShut {NoStop}%
\bibitem [{\citenamefont {D'Alessio}\ and\ \citenamefont
  {Rigol}(2014)}]{DAlessio2014}%
  \BibitemOpen
  \bibfield  {author} {\bibinfo {author} {\bibfnamefont {L.}~\bibnamefont
  {D'Alessio}}\ and\ \bibinfo {author} {\bibfnamefont {M.}~\bibnamefont
  {Rigol}},\ }\href {https://doi.org/10.1103/PhysRevX.4.041048} {\bibfield
  {journal} {\bibinfo  {journal} {Phys. Rev. X}\ }\textbf {\bibinfo {volume}
  {4}},\ \bibinfo {pages} {041048} (\bibinfo {year} {2014})}\BibitemShut
  {NoStop}%
\bibitem [{\citenamefont {Abanin}\ \emph {et~al.}(2015)\citenamefont {Abanin},
  \citenamefont {De~Roeck},\ and\ \citenamefont {Huveneers}}]{Abanin2015}%
  \BibitemOpen
  \bibfield  {author} {\bibinfo {author} {\bibfnamefont {D.~A.}\ \bibnamefont
  {Abanin}}, \bibinfo {author} {\bibfnamefont {W.}~\bibnamefont {De~Roeck}},\
  and\ \bibinfo {author} {\bibfnamefont {F.~m.~c.}\ \bibnamefont {Huveneers}},\
  }\href {https://doi.org/10.1103/PhysRevLett.115.256803} {\bibfield  {journal}
  {\bibinfo  {journal} {Phys. Rev. Lett.}\ }\textbf {\bibinfo {volume} {115}},\
  \bibinfo {pages} {256803} (\bibinfo {year} {2015})}\BibitemShut {NoStop}%
\bibitem [{\citenamefont {Mori}\ \emph {et~al.}(2016)\citenamefont {Mori},
  \citenamefont {Kuwahara},\ and\ \citenamefont {Saito}}]{Mori2016}%
  \BibitemOpen
  \bibfield  {author} {\bibinfo {author} {\bibfnamefont {T.}~\bibnamefont
  {Mori}}, \bibinfo {author} {\bibfnamefont {T.}~\bibnamefont {Kuwahara}},\
  and\ \bibinfo {author} {\bibfnamefont {K.}~\bibnamefont {Saito}},\ }\href
  {https://doi.org/10.1103/PhysRevLett.116.120401} {\bibfield  {journal}
  {\bibinfo  {journal} {Phys. Rev. Lett.}\ }\textbf {\bibinfo {volume} {116}},\
  \bibinfo {pages} {120401} (\bibinfo {year} {2016})}\BibitemShut {NoStop}%
\bibitem [{\citenamefont {Kuwahara}\ \emph {et~al.}(2016)\citenamefont
  {Kuwahara}, \citenamefont {Mori},\ and\ \citenamefont
  {Saito}}]{Kuwahara2016}%
  \BibitemOpen
  \bibfield  {author} {\bibinfo {author} {\bibfnamefont {T.}~\bibnamefont
  {Kuwahara}}, \bibinfo {author} {\bibfnamefont {T.}~\bibnamefont {Mori}},\
  and\ \bibinfo {author} {\bibfnamefont {K.}~\bibnamefont {Saito}},\ }\href
  {https://doi.org/https://doi.org/10.1016/j.aop.2016.01.012} {\bibfield
  {journal} {\bibinfo  {journal} {Annals of Physics}\ }\textbf {\bibinfo
  {volume} {367}},\ \bibinfo {pages} {96} (\bibinfo {year} {2016})}\BibitemShut
  {NoStop}%
\bibitem [{\citenamefont {Mori}\ \emph {et~al.}(2018)\citenamefont {Mori},
  \citenamefont {Ikeda}, \citenamefont {Kaminishi},\ and\ \citenamefont
  {Ueda}}]{Mori:2017qhg}%
  \BibitemOpen
  \bibfield  {author} {\bibinfo {author} {\bibfnamefont {T.}~\bibnamefont
  {Mori}}, \bibinfo {author} {\bibfnamefont {T.~N.}\ \bibnamefont {Ikeda}},
  \bibinfo {author} {\bibfnamefont {E.}~\bibnamefont {Kaminishi}},\ and\
  \bibinfo {author} {\bibfnamefont {M.}~\bibnamefont {Ueda}},\ }\href
  {https://doi.org/10.1088/1361-6455/aabcdf} {\bibfield  {journal} {\bibinfo
  {journal} {J. Phys. B}\ }\textbf {\bibinfo {volume} {51}},\ \bibinfo {pages}
  {112001} (\bibinfo {year} {2018})},\ \Eprint
  {https://arxiv.org/abs/1712.08790} {arXiv:1712.08790 [cond-mat.stat-mech]}
  \BibitemShut {NoStop}%
\bibitem [{\citenamefont {Hayata}\ \emph {et~al.}(2021)\citenamefont {Hayata},
  \citenamefont {Hidaka},\ and\ \citenamefont {Kikuchi}}]{Hayata:2021kcp}%
  \BibitemOpen
  \bibfield  {author} {\bibinfo {author} {\bibfnamefont {T.}~\bibnamefont
  {Hayata}}, \bibinfo {author} {\bibfnamefont {Y.}~\bibnamefont {Hidaka}},\
  and\ \bibinfo {author} {\bibfnamefont {Y.}~\bibnamefont {Kikuchi}},\ }\href
  {https://doi.org/10.1103/PhysRevD.104.074518} {\bibfield  {journal} {\bibinfo
   {journal} {Phys. Rev. D}\ }\textbf {\bibinfo {volume} {104}},\ \bibinfo
  {pages} {074518} (\bibinfo {year} {2021})},\ \Eprint
  {https://arxiv.org/abs/2103.05179} {arXiv:2103.05179 [quant-ph]} \BibitemShut
  {NoStop}%
\bibitem [{\citenamefont {Zache}\ \emph {et~al.}(2023)\citenamefont {Zache},
  \citenamefont {Gonz\'alez-Cuadra},\ and\ \citenamefont
  {Zoller}}]{Zache:2023dko}%
  \BibitemOpen
  \bibfield  {author} {\bibinfo {author} {\bibfnamefont {T.~V.}\ \bibnamefont
  {Zache}}, \bibinfo {author} {\bibfnamefont {D.}~\bibnamefont
  {Gonz\'alez-Cuadra}},\ and\ \bibinfo {author} {\bibfnamefont
  {P.}~\bibnamefont {Zoller}},\ }\href
  {https://doi.org/10.1103/PhysRevLett.131.171902} {\bibfield  {journal}
  {\bibinfo  {journal} {Phys. Rev. Lett.}\ }\textbf {\bibinfo {volume} {131}},\
  \bibinfo {pages} {171902} (\bibinfo {year} {2023})},\ \Eprint
  {https://arxiv.org/abs/2304.02527} {arXiv:2304.02527 [quant-ph]} \BibitemShut
  {NoStop}%
\bibitem [{\citenamefont {Hayata}\ and\ \citenamefont
  {Hidaka}(2023{\natexlab{a}})}]{Hayata:2023puo}%
  \BibitemOpen
  \bibfield  {author} {\bibinfo {author} {\bibfnamefont {T.}~\bibnamefont
  {Hayata}}\ and\ \bibinfo {author} {\bibfnamefont {Y.}~\bibnamefont
  {Hidaka}},\ }\href {https://doi.org/10.1007/JHEP09(2023)126} {\bibfield
  {journal} {\bibinfo  {journal} {JHEP}\ }\textbf {\bibinfo {volume}
  {2023}}\bibfield  {number} {\bibinfo  {number} { (09)},\ \bibinfo {pages}
  {126}},\ }\Eprint {https://arxiv.org/abs/2305.05950} {arXiv:2305.05950
  [hep-lat]} \BibitemShut {NoStop}%
\bibitem [{\citenamefont {Hayata}\ and\ \citenamefont
  {Hidaka}(2023{\natexlab{b}})}]{Hayata:2023bgh}%
  \BibitemOpen
  \bibfield  {author} {\bibinfo {author} {\bibfnamefont {T.}~\bibnamefont
  {Hayata}}\ and\ \bibinfo {author} {\bibfnamefont {Y.}~\bibnamefont
  {Hidaka}},\ }\href {https://doi.org/10.1007/JHEP09(2023)123} {\bibfield
  {journal} {\bibinfo  {journal} {JHEP}\ }\textbf {\bibinfo {volume}
  {2023}}\bibfield  {number} {\bibinfo  {number} { (09)},\ \bibinfo {pages}
  {123}},\ }\Eprint {https://arxiv.org/abs/2306.12324} {arXiv:2306.12324
  [hep-lat]} \BibitemShut {NoStop}%
\bibitem [{\citenamefont {Gils}\ \emph {et~al.}(2009)\citenamefont {Gils},
  \citenamefont {Trebst}, \citenamefont {Kitaev}, \citenamefont {Ludwig},
  \citenamefont {Troyer},\ and\ \citenamefont {Wang}}]{Gils2009}%
  \BibitemOpen
  \bibfield  {author} {\bibinfo {author} {\bibfnamefont {C.}~\bibnamefont
  {Gils}}, \bibinfo {author} {\bibfnamefont {S.}~\bibnamefont {Trebst}},
  \bibinfo {author} {\bibfnamefont {A.}~\bibnamefont {Kitaev}}, \bibinfo
  {author} {\bibfnamefont {A.~W.~W.}\ \bibnamefont {Ludwig}}, \bibinfo {author}
  {\bibfnamefont {M.}~\bibnamefont {Troyer}},\ and\ \bibinfo {author}
  {\bibfnamefont {Z.}~\bibnamefont {Wang}},\ }\href
  {https://doi.org/10.1038/nphys1396} {\bibfield  {journal} {\bibinfo
  {journal} {Nature Physics}\ }\textbf {\bibinfo {volume} {5}},\ \bibinfo
  {pages} {834} (\bibinfo {year} {2009})}\BibitemShut {NoStop}%
\bibitem [{\citenamefont {Rodríguez‐Plaza}(1991)}]{10.1063/1.529497}%
  \BibitemOpen
  \bibfield  {author} {\bibinfo {author} {\bibfnamefont {M.~J.}\ \bibnamefont
  {Rodríguez‐Plaza}},\ }\href {https://doi.org/10.1063/1.529497} {\bibfield
  {journal} {\bibinfo  {journal} {Journal of Mathematical Physics}\ }\textbf
  {\bibinfo {volume} {32}},\ \bibinfo {pages} {2020} (\bibinfo {year}
  {1991})},\ \Eprint
  {https://arxiv.org/abs/https://pubs.aip.org/aip/jmp/article-pdf/32/8/2020/19212819/2020\_1\_online.pdf}
  {https://pubs.aip.org/aip/jmp/article-pdf/32/8/2020/19212819/2020\_1\_online.pdf}
  \BibitemShut {NoStop}%
\bibitem [{\citenamefont {Javadi-Abhari}\ \emph {et~al.}(2024)\citenamefont
  {Javadi-Abhari}, \citenamefont {Treinish}, \citenamefont {Krsulich},
  \citenamefont {Wood}, \citenamefont {Lishman}, \citenamefont {Gacon},
  \citenamefont {Martiel}, \citenamefont {Nation}, \citenamefont {Bishop},
  \citenamefont {Cross}, \citenamefont {Johnson},\ and\ \citenamefont
  {Gambetta}}]{qiskit_paper}%
  \BibitemOpen
  \bibfield  {author} {\bibinfo {author} {\bibfnamefont {A.}~\bibnamefont
  {Javadi-Abhari}}, \bibinfo {author} {\bibfnamefont {M.}~\bibnamefont
  {Treinish}}, \bibinfo {author} {\bibfnamefont {K.}~\bibnamefont {Krsulich}},
  \bibinfo {author} {\bibfnamefont {C.~J.}\ \bibnamefont {Wood}}, \bibinfo
  {author} {\bibfnamefont {J.}~\bibnamefont {Lishman}}, \bibinfo {author}
  {\bibfnamefont {J.}~\bibnamefont {Gacon}}, \bibinfo {author} {\bibfnamefont
  {S.}~\bibnamefont {Martiel}}, \bibinfo {author} {\bibfnamefont {P.~D.}\
  \bibnamefont {Nation}}, \bibinfo {author} {\bibfnamefont {L.~S.}\
  \bibnamefont {Bishop}}, \bibinfo {author} {\bibfnamefont {A.~W.}\
  \bibnamefont {Cross}}, \bibinfo {author} {\bibfnamefont {B.~R.}\ \bibnamefont
  {Johnson}},\ and\ \bibinfo {author} {\bibfnamefont {J.~M.}\ \bibnamefont
  {Gambetta}},\ }\href {https://arxiv.org/abs/2405.08810} {\bibinfo {title}
  {Quantum computing with qiskit}} (\bibinfo {year} {2024}),\ \Eprint
  {https://arxiv.org/abs/2405.08810} {arXiv:2405.08810 [quant-ph]} \BibitemShut
  {NoStop}%
\bibitem [{\citenamefont {Kim}\ \emph {et~al.}(2023)\citenamefont {Kim},
  \citenamefont {Eddins}, \citenamefont {Anand}, \citenamefont {Wei},
  \citenamefont {van~den Berg}, \citenamefont {Rosenblatt}, \citenamefont
  {Nayfeh}, \citenamefont {Wu}, \citenamefont {Zaletel}, \citenamefont
  {Temme},\ and\ \citenamefont {Kandala}}]{Kim2023}%
  \BibitemOpen
  \bibfield  {author} {\bibinfo {author} {\bibfnamefont {Y.}~\bibnamefont
  {Kim}}, \bibinfo {author} {\bibfnamefont {A.}~\bibnamefont {Eddins}},
  \bibinfo {author} {\bibfnamefont {S.}~\bibnamefont {Anand}}, \bibinfo
  {author} {\bibfnamefont {K.~X.}\ \bibnamefont {Wei}}, \bibinfo {author}
  {\bibfnamefont {E.}~\bibnamefont {van~den Berg}}, \bibinfo {author}
  {\bibfnamefont {S.}~\bibnamefont {Rosenblatt}}, \bibinfo {author}
  {\bibfnamefont {H.}~\bibnamefont {Nayfeh}}, \bibinfo {author} {\bibfnamefont
  {Y.}~\bibnamefont {Wu}}, \bibinfo {author} {\bibfnamefont {M.}~\bibnamefont
  {Zaletel}}, \bibinfo {author} {\bibfnamefont {K.}~\bibnamefont {Temme}},\
  and\ \bibinfo {author} {\bibfnamefont {A.}~\bibnamefont {Kandala}},\ }\href
  {https://doi.org/10.1038/s41586-023-06096-3} {\bibfield  {journal} {\bibinfo
  {journal} {Nature}\ }\textbf {\bibinfo {volume} {618}},\ \bibinfo {pages}
  {500} (\bibinfo {year} {2023})}\BibitemShut {NoStop}%
\bibitem [{\citenamefont {Fishman}\ \emph {et~al.}(2022)\citenamefont
  {Fishman}, \citenamefont {White},\ and\ \citenamefont
  {Stoudenmire}}]{itensor}%
  \BibitemOpen
  \bibfield  {author} {\bibinfo {author} {\bibfnamefont {M.}~\bibnamefont
  {Fishman}}, \bibinfo {author} {\bibfnamefont {S.~R.}\ \bibnamefont {White}},\
  and\ \bibinfo {author} {\bibfnamefont {E.~M.}\ \bibnamefont {Stoudenmire}},\
  }\href {https://doi.org/10.21468/SciPostPhysCodeb.4} {\bibfield  {journal}
  {\bibinfo  {journal} {SciPost Phys. Codebases}\ ,\ \bibinfo {pages} {4}}
  (\bibinfo {year} {2022})}\BibitemShut {NoStop}%
\bibitem [{\citenamefont {Kechedzhi}\ \emph {et~al.}(2024)\citenamefont
  {Kechedzhi}, \citenamefont {Isakov}, \citenamefont {Mandr\`a}, \citenamefont
  {Villalonga}, \citenamefont {Mi}, \citenamefont {Boixo},\ and\ \citenamefont
  {Smelyanskiy}}]{Kechedzhi:2023swt}%
  \BibitemOpen
  \bibfield  {author} {\bibinfo {author} {\bibfnamefont {K.}~\bibnamefont
  {Kechedzhi}}, \bibinfo {author} {\bibfnamefont {S.~V.}\ \bibnamefont
  {Isakov}}, \bibinfo {author} {\bibfnamefont {S.}~\bibnamefont {Mandr\`a}},
  \bibinfo {author} {\bibfnamefont {B.}~\bibnamefont {Villalonga}}, \bibinfo
  {author} {\bibfnamefont {X.}~\bibnamefont {Mi}}, \bibinfo {author}
  {\bibfnamefont {S.}~\bibnamefont {Boixo}},\ and\ \bibinfo {author}
  {\bibfnamefont {V.}~\bibnamefont {Smelyanskiy}},\ }\href
  {https://doi.org/10.1016/j.future.2023.12.002} {\bibfield  {journal}
  {\bibinfo  {journal} {Future Gener. Comput. Syst.}\ }\textbf {\bibinfo
  {volume} {153}},\ \bibinfo {pages} {431} (\bibinfo {year} {2024})},\ \Eprint
  {https://arxiv.org/abs/2306.15970} {arXiv:2306.15970 [quant-ph]} \BibitemShut
  {NoStop}%
\bibitem [{\citenamefont {Sivarajah}\ \emph {et~al.}(2020)\citenamefont
  {Sivarajah}, \citenamefont {Dilkes}, \citenamefont {Cowtan}, \citenamefont
  {Simmons}, \citenamefont {Edgington},\ and\ \citenamefont
  {Duncan}}]{tket_paper}%
  \BibitemOpen
  \bibfield  {author} {\bibinfo {author} {\bibfnamefont {S.}~\bibnamefont
  {Sivarajah}}, \bibinfo {author} {\bibfnamefont {S.}~\bibnamefont {Dilkes}},
  \bibinfo {author} {\bibfnamefont {A.}~\bibnamefont {Cowtan}}, \bibinfo
  {author} {\bibfnamefont {W.}~\bibnamefont {Simmons}}, \bibinfo {author}
  {\bibfnamefont {A.}~\bibnamefont {Edgington}},\ and\ \bibinfo {author}
  {\bibfnamefont {R.}~\bibnamefont {Duncan}},\ }\href
  {https://doi.org/10.1088/2058-9565/ab8e92} {\bibfield  {journal} {\bibinfo
  {journal} {Quantum Science and Technology}\ }\textbf {\bibinfo {volume}
  {6}},\ \bibinfo {pages} {014003} (\bibinfo {year} {2020})}\BibitemShut
  {NoStop}%
\end{thebibliography}%
\end{document}